**Optoionic Impedance Spectroscopy (OIS): a model-less technique for in-situ electrochemical characterization of mixed ionic electronic conductors**

*Paul Nizet, Francesco Chiabrera*, Yunqing Tang, Nerea Alayo, Beatrice Laurenti, Federico Baiutti, Alex Morata, Albert Tarancón**

Paul Nizet, Francesco Chiabrera, Nerea Alayo, Beatrice Laurenti, Federico Baiutti, Alex Morata, Albert Tarancón
Department of Advanced Materials for Energy Applications, Catalonia Institute for Energy Research (IREC), Jardins de les Dones de Negre 1, 08930, Sant Adrià del Besòs, Barcelona, Spain
E-mail: fchiabrera@irec.cat, atarancon@irec.cat

Yunqing Tang
College of Aerospace and Civil Engineering, Harbin Engineering University, Nantong Street 145, 150001, Harbin, P. R. China

Albert Tarancón
Catalan Institution for Research and Advanced Studies (ICREA), Passeig Lluís Companys 23, 08010, Barcelona, Spain

Keywords: mixed ionic electronic conductor (MIEC), in-situ, electrochemical ionic insertion, chemical capacitance, optochemical, La1-xSrxFeO3 (LSF), ionic gating

Functional properties of mixed ionic electronic conductors (MIECs) can be radically modified by (de)insertion of mobile charged defects. A complete control of this dynamic behaviour has multiple applications in a myriad of fields including advanced computing, data processing, sensing or energy conversion. However, the effect of different MIEC's state-of-charge is not fully understood yet and there is a lack of strategies for fully controlling the defect content in a material. In this work we present a model-less technique to characterize ionic defect concentration and ionic insertion kinetics in MIEC materials: Optoionic Impedance Spectroscopy (OIS). The proof of concept and advantages of OIS are demonstrated by studying the oxygen (de)insertion in thin films of hole-doped perovskite oxides. Ion migration into/out of the studied materials is achieved by the application of an electrochemical potential, achieving stable and reversible modification of its optical properties. By tracking the dynamic variation of optical properties depending on the gating conditions, OIS enables to extract electrochemical parameters involved in the electrochromic process. The results demonstrate the capability of the technique to effectively characterize the kinetics of single- and even multi-layer systems. The technique can be employed for studying underlying mechanisms of the response characteristics of MIEC-based devices.

## 1. Introduction

Mixed ionic electronic conductors (MIECs) are an important class of materials characterized by simultaneous conduction of ions and electrons. One key aspect of this family of materials is the ability to host different types of mobile charged point defects, such as vacancies, interstitials, or substitutional ions[1]. The application of an electrochemical potential to a MIEC using an ion conducting electrolyte can modify the chemical equilibrium of point defects in MIEC materials, forcing ion insertion/extraction until the equilibrium is restored. This fundamental characteristic has allowed the development of non-stoichiometric materials such as efficient insertion electrodes for lithium-ion batteries or solid oxide cells' electrodes for hydrogen technology[2–5]. Additionally, the strong correlation between ionic point defects and structural/electronic properties in MIECs offers a useful tool to modulate their functional properties.[6–9] For instance, in transition metal oxides, the concentration of oxygen vacancies strongly affects the electronic conductivity, magnetic order or optical properties.[10–12] For these reason, the modulability of MIEC properties through the control of the concentration of point defects holds potential for application in novel and advanced devices such as neuromorphic transistors, memristors, and light modulators[13,14], where the time-response, reproducibility, and precise control of the functional layer properties are key.[15,16]

In all these systems, it is of high importance to understand equilibrium ionic defect concentration and kinetics of ionic insertion, as well as their effect on a specific functionality, possibly in-situ and/or operando. Among all the techniques that have been developed to determine the modulability of a system, Impedance Spectroscopy (IS) is one of the most used. IS belongs to a family of methodologies involving the application of an external sinusoidal stimulus, spanning a wide range of frequencies, to a system in equilibrium or in steady state, with measurement of the sinusoidal response. Among the various IS techniques, Electrochemical Impedance Spectroscopy (EIS), where both the stimulus and the measured response are electrical, is the most used.[17] The importance of EIS over other electrochemical techniques lies in its ability to discriminate among various electrical, electrochemical, and physical resistive processes that take place during operation of real electrochemical systems. For instance, EIS has been used for corrosion studies, semiconductor science and technology, solid-oxide cells and battery characterization.[18] The objective of these analysis is to extract different time constants $\tau$ of the systems. Each of those $\tau$ is related to the time behavior of a process involving electron and/or ionic migration or reaction.

Beyond traditional IS, non-electrical IS (nEIS), based on the use of a non-electrical stimulus or the measurement of a non-electrical response, presents a number of potential advantages especially for the investigation of MIEC materials and devices. As an example, Swallow et al. investigated the oxygen-vacancy induced chemical expansion of Pr-doped ceria thin films by nanoscale electrochemomechanical spectroscopy (NECS), a technique based on the detection of the sample's deflection under the application of a sinusoidal electrochemical potential.[19] NECS allowed to fully characterize the dynamics of the chemical expansion, offering important insights into the controlling kinetics of high-temperature electrochemical actuators. One notable example of nEIS based on a non-electrical stimulus is the work of Defferriere et al.[20]**,** where the authors developed intensity-modulated photocurrent spectroscopy (IMPS) to get insights into the increase of ionic conductivity of Gd-doped ceria thin films when excited by light modulation in the frequency domain. Thanks to the ability of IMPS and traditional EIS to separate resistive ionic contributions, they concluded that the observed drastic increase of ionic conductivity originated by a decrease of grain boundary resistance.

Interestingly, nEIS approaches can be employed to get information on processes that take place following ionic insertion. In this case, it is necessary to find a measurable functional property that varies when the ionic concentration of a defect species is modified. For instance, it is well known that optical properties such as color[21], optical transmission[22–24] or optical conductivity [25,26] are very sensitive to the defect concentration. An interesting approach is the one made by Manka *et al.*[27] Their research work introduced a novel technique for characterizing Li-ion battery electrodes based on quantifying optical transmission changes. This innovative approach, defined as optical impedance, enabled the measurement of optical property variations as a response to an applied sinusoidal potential. However, the use of a transmission mode involves an important drawback, as the substrate must be light-transparent. This limitation restricts the technique's application to specific cases or to limited photon energy windows. A different approach to study optical properties changes with ion concentrations is the use of in-situ Spectroscopic Ellipsometry (SE), an optical non-destructive technique sensitive to various structural and optical attributes of a thin film, including optical constants and thickness, regardless of the substrate's optical properties. This strategy was adopted for the study of Li content, thickness, and Li diffusivity in Li-cathode materials for solid state batteries.[28–31] Moreover, In-situ SE permitted the investigation of the defect chemistry of perovskite oxides at intermediate temperatures on solid state electrolytes [25] and at room temperature in alkaline liquid electrolyte [26]. To do so, these studies employ a model to determine the optical properties of each material composing the sample, which can add severe complications in the case of multilayers or novel materials.[32] A study by Lazovski et al. also took advantage of In-situ SE signal sensitivity to changes in optical properties to study oxygen diffusion in electrolyte ionic conductor films.[33,34] A nEIS approach was employed by applying an AC potential across the layer. They could observe a variation of optical constants, which they assigned to an electric-field induced accumulation of ionic defects in purely ionic conductors. Although their work demonstrates the great sensitivity of ellipsometry to detect point defects, DC analysis was finally needed to measure the activation energy of oxygen diffusion.

The advantages and constraints of the aforementioned methods highlight the interest in developing a nEIS based on in-situ SE technique that does not require the use of optical models. This paper presents optoionic impedance spectroscopy (OIS), a nEIS technique based on accessible and flexible optical methods for characterizing the ionic insertion and diffusion in MIEC thin films using in-situ SE. OIS is based on the measurement of the optical response upon application of an oscillatory AC electrochemical potential to a MIEC/electrolyte system. First, we set the theoretical basis of the technique by deriving the general expression of the OIS response of a 1D MIEC material distributed element and comparing it with the well-known EIS case. Then, four different case studies are carried out for the demonstration of OIS technique capabilities: (i) Validation of the technique through the measurement of the chemical capacitance response of $La_{0.5}Sr_{0.5}FeO_{3-\delta}$ (LSF50) thin films as a function of electrochemical bias. Based on the sensitive response of the ellipsometry parameters of $La_{1-x}Sr_xFeO_{3-\delta}$ (LSFx) thin films to Nernstian voltages[25], we demonstrated the capability of OIS of retrieving the chemical capacitance without employing any optical fitting model. (ii) Application of the OIS technique for the characterization of the chemical capacitance of $SrFeO_{3-\delta}$ (SFO) thin films, a material with less established defect chemistry knowledge and characterized by a topotactic phase transition under reducing conditions. (iii) Application of the OIS technique for the characterization of chemical capacitance of complex bilayer LSF50/SFO. The aim of this part is to show that the technique is able to measure optical changes coming from complex structures. (iv) Application of OIS is extended to investigate the oxygen diffusion of LSF50 at room temperature in liquid alkaline electrolyte.

## 2. Methods and theory

### 2.1 In-situ spectroscopy ellipsometry

As an optical technique, Spectroscopic Ellipsometry (SE) relies on the measurement of a light beam after the interaction with the sample. A polarized light beam enters the stage at a precisely defined angle of incidence and is focused on the desired spot. After going through the sample, the light beam is reflected and analyzed. A detector measures the polarization variation (from linear to elliptical) of the reflected light. The complex reflectance is described by two main parameters, namely the relative variation of amplitude ($\Psi$) and phase ($\Delta$) of the polarized light as a function of the photon energy of the incoming light (see **Figure 1a and 1b**). Ellipsometry allows for the determination of different properties of a thin film, such as thickness, roughness, refractive (n) and extinction (k) coefficients[35]. In a single semi-infinite material, the optical parameters of the material are directly related to the reflected light, i.e. higher n (k) determines a decrease (increase) of $\Psi$ ($\Delta$). However, when studying multilayers, the relation between $\Psi$ and $\Delta$ and optical properties is not straightforward, as light reflects at each interface creating extra interferences in the measured spectra. For complex samples (e.g. multilayers), a model-based analysis approach relying on a fitting procedure is needed in order to extract optical information of each component from the ellipsometry dispersion parameters $\Delta$ and $\Psi$. [32]

In this context, Figure 1 exemplifies the main concept behind in-situ SE for the study of electrochemical ionic insertion, where the optical properties of a thin film are investigated while inducing ionic insertion (extraction) from a solid/liquid electrolyte through the application of an electrochemical potential: as ions enter (leave) a MIEC thin film, a variation of the oxidation state or electrons concentration is induced in the material as a charge compensation mechanism, typically promoting a large variation of electronic structure (**Figure 1c**) and hence of optical reflectance of a sample (Figure 1b). In Figure 1b, as a matter of example, the optical properties have been simulated using a single Lorentzian oscillator, with decreasing bandgap as ions are inserted in the structure. Alternatively, variation of optical properties and SE spectra can be related to changes of thickness, roughness, or formation of space charge during ion insertion [36,37]. In this way, a relation between optical properties and concentration of ionic defects can be obtained to study electrochemical properties of MIEC thin films by SE (see **Figure 1d**).[25,26,28–31] Usually, the use of a model to determine the optical properties is needed and the final relation may present a non-linear behavior over the entire insertion range, which complicates the interpretation and quantification of the ionic content by SE (see Figure 1d).

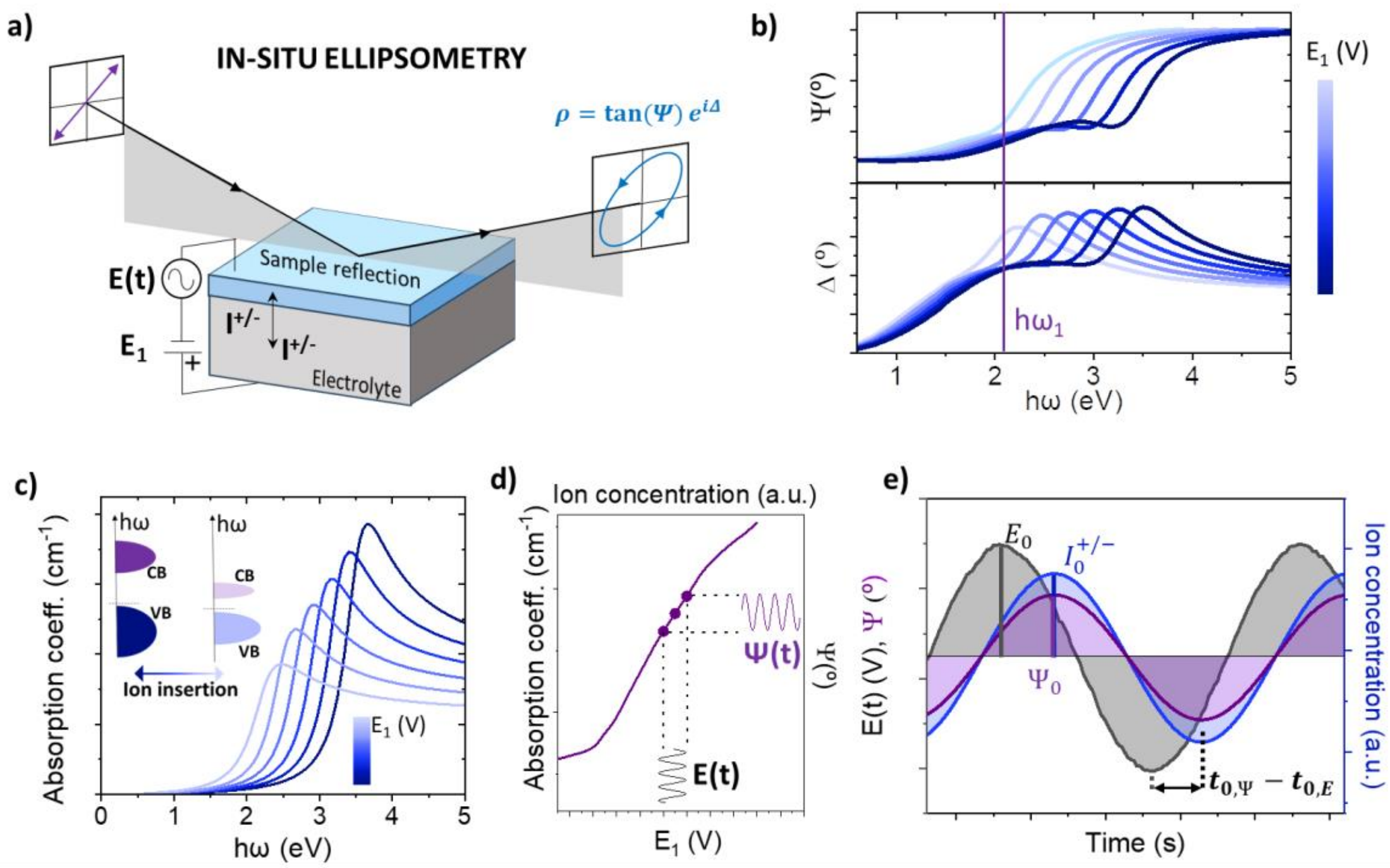

**Figure 1**. a) Configuration of in-situ SE: the complex reflectance is measured while applying a DC or AC electrochemical potential and forcing ion insertion from a solid or liquid electrolyte. b) Exemplary SE amplitude (Ψ) and phase (Δ) spectra simulated for a thin film presenting a variation of absorption upon ion insertion shown in c). The inset shows the corresponding simplified band structure upon ionic insertion. d) A relation between ion concentration and optical absorption can be found but not necessarily linear over the entire range. The application of a small sinusoidal signal will instead produce a linear response, detected by measuring the variation of SE dispersion parameters (e.g. Ψ). e) Schematic of the sinusoidal inputs and optical outputs elucidating the OIS concept.

## 2.2 Optoionic Impedance Spectroscopy (OIS)

As an impedance measurement, OIS consists in the implementation of an alternate input signal with variable frequency and the simultaneous read out of the corresponding output signal. The alternate input stimulus is an electrochemical potential in the form a sinusoidal wave:

$$E(t) = E_0 \sin(\omega t) + E_1 \quad (1)$$

Where $E_1$ is the DC electrochemical bias, $E_0$ indicates the amplitude of the sinusoidal-shape wave and $\omega$ is the angular frequency.

As output signal, OIS directly considers the complex reflectance, either defined by the angle Ψ, Δ or a combination of those (e.g. complex module). Here, Ψ was chosen because it showed a better signal strength, however the use of Δ is analogous and may be preferred under specific conditions. Due to the relation between ionic concentration and optical properties described in the previous section, if the electrical potential (see Equation (1)) induces ionic insertion in the sample, the optical dispersion parameters Δ, Ψ will also change. It is essential to acknowledge that due to the complexity of signal interpretation, especially when having multiple-layer samples, a direct linear correlation between the external stimulus and optical output may not always exist. Nevertheless, for sufficiently small voltage differences (ΔV), specific segments can be approximated as linear, see Figure 1d. In OIS, we assume linearity in analogy with EIS. This is confirmed at the beginning of each experiment following the procedure explained in Experimental section. A more in-depth discussion on the linearity relation and the strategy used for ensuring it is presented in Supporting Information Section S1. Consequently, if a sinusoidal voltage is applied as the external excitation also the signal read by the ellipsometer would be a sine wave with the same frequency (**Figure 1e**):

$$\Psi(t) = \Psi_0(\omega) \sin\big(\omega t + \varphi(\omega)\big) + \Psi_1 \quad (2)$$

Where $\Psi_1$ is the constant background signal, $\Psi_0$ indicates the amplitude of the sinusoidal-shape wave and $\varphi(\omega)$ represents the phase lag between the optical and the electrical signals. Similar to the mechanism of EIS and electrochemomechanical spectroscopy presented by Swallow et al. for the material $Pr_xCe_{1-x}O_{2-\delta}$ [19] the optical admittance can be defined as the output signal over the input signal:

$$Y_{OIS}(\omega) = \frac{\Psi_0(\omega)}{E_0(\omega)} \Big(\sin\big(\varphi(\omega)\big) + j \cos\big(\varphi(\omega)\big)\Big) \quad (3)$$

Note that in the above equations (2) and (3) both the amplitude and the phase lag $\varphi$ have a frequency dependence.

### *2.2.1 Generalized OIS response*

To better understand the formulism of the optical admittance, a straightforward parallelism can be done with its electrical counterpart:

$$Y_{EIS}(\omega) = \frac{I(\omega)}{E(\omega)} = \frac{I(\omega)}{I(\omega) \cdot Z_T} = \frac{1}{Z_T} \quad (4)$$

Where $Z_T$ stands for the impedance of the circuit. In the case of OIS, the observable parameter is the variation of complex reflectance in the sample, represented in this work by the response $\Psi$. As previously explained, this optical response may be generated by different phenomena (e.g. sample expansion and change of optical constants) that, for small perturbation signals, vary linearly with the ion concentration of the intercalating species. Therefore, the OIS admittance can be directly related to the amount of inserted/extracted charged ions $q_{mat}(\omega)$, as:

$$Y_{OIS} = \frac{\Psi(\omega)}{E(\omega)} = \frac{E_{mat}(\omega)}{E(\omega)} \cdot \frac{q_{mat}(\omega)}{E_{mat}(\omega)} \cdot \frac{\Psi(\omega)}{q_{mat}(\omega)} = H(\omega) \cdot \frac{q_{mat}(\omega)}{E_{mat}(\omega)} . N \quad (5)$$

where $E_{mat}(\omega)$ stands for the voltage drop across the studied material. The resulting expression presents the product of three elements: the transfer function of the element under study $H(\omega)$ (*i.e.* the ratio between the voltage drop in the studied film and the potential applied to the whole circuit), the ratio $q_{mat}(\omega)/E_{mat}(\omega)$ and, the ratio between an optical parameter and the ionic charge, which can be reduced to a constant, N, through the linearity assumption. The transfer function can be easily extracted when solving the equivalent electrical circuit of the system under study. Therefore, the interpretation of OIS spectra simplifies into the evaluation of $\frac{q_{mat}(\omega)}{E_{mat}(\omega)}$, namely the relative variation of ion concentration of a MIEC layer under the application of an AC bias. This is the main difference between EIS and OIS: While EIS measures the variation of charge in time (*i.e.* current), OIS probes the chemical charge per unit potential, *viz* the capacitive contribution resulting from the accumulating ions. It is important to note that this charge is only the one that accumulates in the layer due to a battery-like response, charge flowing through the system would not be measured by OIS.

### *2.2.2 OIS of finite-space diffusion in 1D MIEC*

In this section we derive the generalized behavior of the $Y_{OIS}$ of a 1D MIEC when being ionically (dis)charged via voltage application. To simulate the behavior of different type of MIECs, the general transmission line is generally adopted [18,38]. In this element, the MIEC is described by two parallel rails for ions and electrons, respectively.

In this work, we restricted the model to the characteristic finite-space diffusion case and that enables the ion (electrons) insertion through one of its sides, i.e., with a blocking electrode at the opposite end (see **Figure 2a**). $R_i = \sum dR_i$ will be taken as the ionic resistance (the electronic rail is short-circuited due to high electronic conductivity of the MIEC) and $C = \sum dC$ as the ionic capacity of the material. We will consider R as the resistance to the ion insertion/extraction at the material's interface. This resistive contribution may also present a contribution of other resistive elements placed in series with the layer, e.g., electrolyte resistance. The frequency response of the system will strongly depend on the $R_i/R$ ratio as well as the value of $C$. First, we will evaluate its effect in the EIS. The impedance element of a MIEC with these boundary conditions is known as *Warburg open* and its expression is:

$$Z_{W_o} = \frac{R_i}{X}\coth(X) \qquad where\ X = \sqrt{i\omega R_i C} \tag{6}$$

Which tends to $Z_C = {}^{1}/_{i\omega C}$ for low $R_i$ values. **Figure 2c-e** shows the EIS Bode and Nyquist plots for a system with fixed $C$ and $R$, but different values of $R_i$. When the diffusion does not limit the process, i.e. a low $R_i$ value, a vertical line is shown in the Nyquist plot at Re(Z)=R, which corresponds to an increase in the diphase as the frequency is reduced. The circuit therefore reduces to a single RC element, where the ions entering in the MIEC charge the chemical capacitance with a characteristic frequency equal to $f = \frac{1}{2\pi RC}$ (see **Figure 2b**). When increasing the ionic resistivity, the charging of the capacitance happens at a slower rate and diffusion becomes non-negligible. This gives rise to the characteristic 45° slope in the Nyquist impedance plot before reaching its characteristic frequency $f = \frac{1}{2\pi R_i C}$, in which the system recovers the vertical line in the Nyquist plot but at a higher Re(Z) value.

For the OIS response, the different elements composing Equation 5 must be derived. First, the transfer function can be obtained as

$$H(\omega) = \frac{E_{MIEC}(\omega)}{E(\omega)} = \frac{I(\omega)\cdot Z_{W_o}}{I(\omega)\cdot(Z_{W_o}+R)} = \frac{1}{1+\mathrm{R}/Z_{W_o}} \tag{7}$$

The second important term to solve to get the OIS expression that fits each model system is the $\frac{q_{mat}(\omega)}{E_{mat}(\omega)}$ relation. For the transmission line shown in Figure 2a, the total accumulated charge is distributed along the different capacities, which can be seen as the different charging state of different points along the thickness of the material. The closer to the ionic insertion region, the faster the capacity will charge, as ions must cross the different $dR_i$ to reach the inner part of the material. Electrically, the state of charge of each capacitor of the transmission line would be related to the voltage drop at its ends as $dq(x) = dC \cdot (E_i(x) - E_e(x))$, where $E_i(x)$ and $E_e(x)$ are the voltage in the ionic and electronic path, respectively, and $dC = \frac{C}{L} dx$. Since no electrical losses are considered in the electronic pathway, the accumulated charge only depends on the potential drop of the ionic rail, $E_i(x)$. The total charge per unit potential can be therefore obtained by integration over the layer thickness, as:

$$\frac{q_{mat}(\omega)}{E_{mat}(\omega)} = \frac{\int_0^L C/L \cdot E_i(x) dx}{E_{mat}(\omega)} \tag{8}$$

By considering the set of differential equations and boundary conditions presented in Supporting Information Section S2**,** the equation leads to the following expression:

$$\frac{q_{mat}(\omega)}{E_{mat}(\omega)} = C \cdot \frac{tanh(X)}{X} \quad where\ X = \sqrt{i\omega R_i C} \tag{9}$$

The derivation of $\frac{q_{mat}(\omega)}{V_{mat}(\omega)}$ for more complex systems is presented in Supporting Information Section S2. Therefore, the solution of OIS admittance of finite-space diffusion in 1D MIEC can be obtained by substituting Equations 7 and 9 in Equation 5, as:

$$Y_{OIS} = \frac{C \cdot N}{\mathrm{X} \cdot \coth(X) + i\omega RC} \tag{10}$$

Equation 10 clearly shows that the OIS has the formal shape of a capacitance. To better visualize the effect of the different parameters on $Y_{OIS}$, **Figures 2f-h** show the normalized OIS admittance Bode and Nyquist plots for different ionic diffusions (by dividing $Y_{OIS}$ by $C \cdot N$). For the case with a low ionic resistance $R_i$, the expression reduces to the one of an RC circuit. At a given high frequency, the kinetic limitations prevent to reach the equilibrium at the applied voltage, resulting in a small variation of charge and a large time delay, i.e. a low $|Y_{OIS}(\omega)|$ value and an increase in the phase lag $\varphi(\omega)$ value. Lowering the frequency, the system has more time to reach the equilibrium, resulting in a higher $|Y_{OIS}(\omega)|$ and a smaller delay. It is clearly seen that as the diffusivity is lowered (higher $R_i$) and becomes limiting, the response at higher frequencies starts to change its shape, as the defects do not have time to charge the film uniformly and only a portion of the material (closer to the electrode) is charged. Another way to see this phenomenon is thinking on the effective capacity being charged. As the frequency is increased, a lesser percentage of the film can reach the equilibrium (de)intercalated state and so the effective capacity being measured is lowered making the diphase lower as well. For sufficiently high $R_i$ values, a plateau at 45º is seen, corresponding to the same 45º observed in the EIS. The effect on the Nyquist plot is also evident, as the shape for the RC circuit corresponds to a perfect semicircle and the one for the diffusion limited case presents an initial 45º slope prior a smaller semicircle. It is important to note that at low frequencies, all the spectrum ends up in Re(Z)=1 due to the $Y_{OIS}(\omega)$ normalization. The Ri/R ratio is indeed a key parameter to determine if the system is in the diffusion-limited range or not, as will be shown in the experimental examples. Importantly, the theoretical framework introduced here for OIS of MIEC materials can be easily extended to other techniques sensitive to the ionic charge of a material, such as NECS.[19]

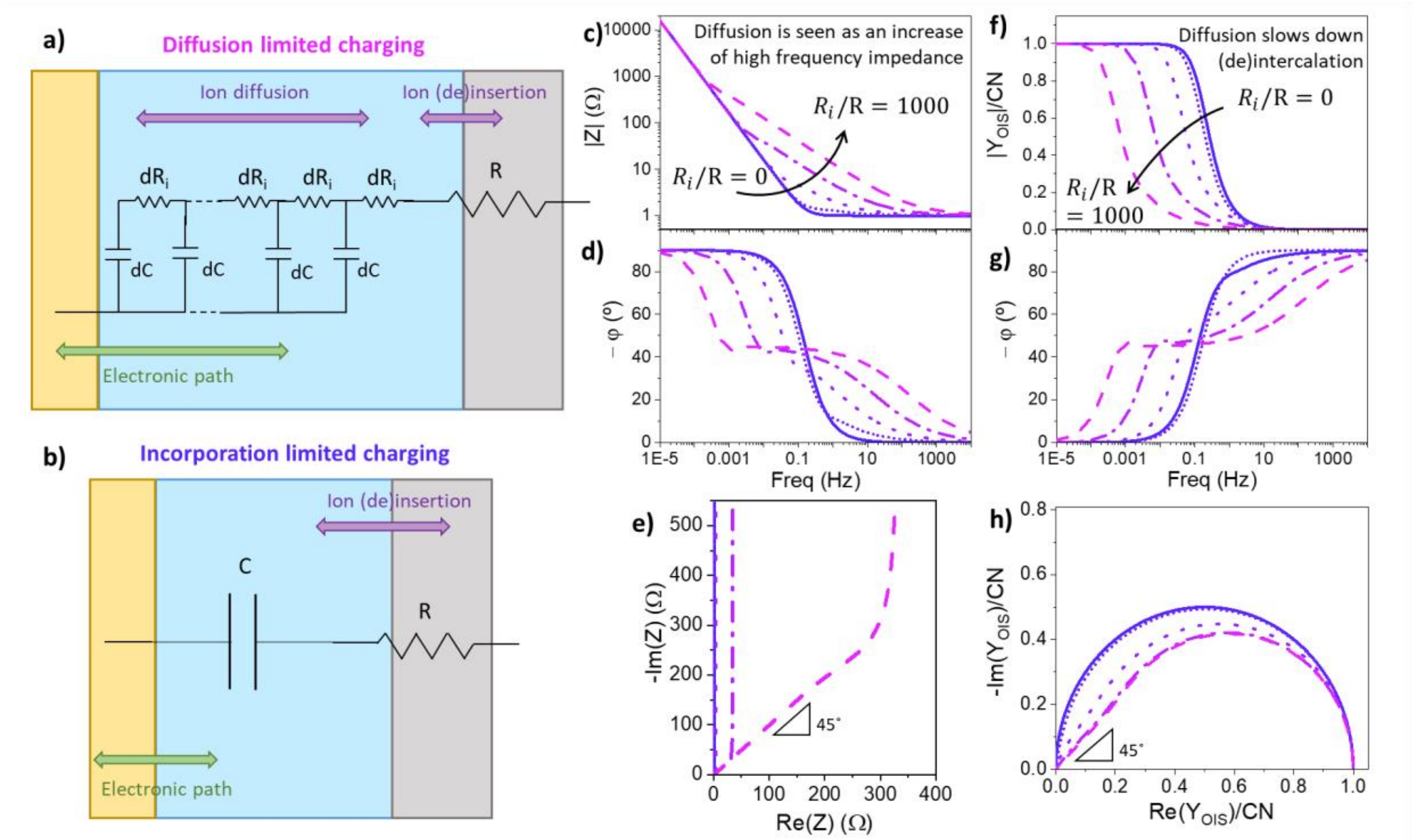

**Figure 2**. a) Equivalent circuit for the diffusion-limited ion (de)insertion in a MIEC. b) Simplification of the circuit for the case of neglectable ionic diffusion. c-d) EIS Bode plots and e) Nyquist plot for circuit (a) considering different R/Ri values. The OIS Admittance plots for the same $R_i/R$ values are also presented: f-g) Shows the OIS Bode Admittance plots and h) the Nyquist Admittance plot

*2.2.3 OIS for oxygen-ion insertion in MIEC thin films*

As commented in the introduction, all the different case studies presented in this work concern the measurement of oxygen-ion electrochemical properties of different single-/composites LSFx thin films. DC and/or AC electrochemical potential was applied through the sample while measuring the optical response by SE. The use of an electrolyte allows to directly probe different regions of the Brouwer diagram through voltage-driven insertion/extraction of oxygen ions, according to the Nerst potential: [25,39]

$$pO_{2,eq} = pO_2 \cdot \exp\left(\frac{4eE}{k_b T}\right) \qquad (11)$$

Where $pO_2, e, k_b$ and T are the oxygen partial pressure, electron charge, Boltzmann constant and temperature, respectively. $E$ is the overpotential on the working electrode. In this sense, the application of an electrochemical potential has the effect of shifting the electrochemical equilibrium of oxygen ions, thus providing the possibility of varying the concentration of point defects through the equilibrium: [25,40]

$$\frac{1}{2}\mathrm{O}_2 + \mathrm{V}_\mathrm{O}^{\cdot\cdot} + 2\mathrm{Fe}_\mathrm{Fe}^\mathrm{x} \leftrightarrow \mathrm{O}_\mathrm{O}^\mathrm{x} + 2\mathrm{Fe}_\mathrm{Fe}^{\cdot} \qquad (12)$$

With point defects written according to Kröger-Vink notation. The application of an electrochemical potential to the system generates an oxygen current flow until the equilibrium between point defects and $pO_{2,eq}$ is reestablished. In fact, this change in the electrochemical equilibrium of oxygen ions is responsible for the capacitive behavior of the LSFx film. The chemical capacitance, understood as the volumetric capacity of the film, is therefore defined as the variation of the oxygen concentration in the layer when the oxygen equilibrium of Equation 12 is electrochemically modified: [25,39]

$$C_{chem} = \frac{e^2}{k_b T} n_{uc} \frac{dc_o}{d\mu_o} \qquad (13)$$

Where $\mu_o$ represents the oxygen chemical potential, and $c_o$ is the concentration of oxygen. In a previous work, we have used this strategy to study the evolution of optical properties of LSFx thin films by in-situ ellipsometry.[25] There, we found a strong relation between electron holes and optical absorption, which was used to track shift from the oxygen equilibrium reaction while applying DC electrochemical potentials. In the present work, we will take advantage of the relation between optical properties and electron holes in LSFx for demonstrating the capabilities of OIS. However, it is important to remark that it is not necessary to understand the origin of these optical changes to perform OIS. As long as any optical-sensitive property (dielectric constant, thickness, roughness) of the film is modified by ion insertion and the applied stimulus is small enough, we can use the linearity stimulus-signal explained in the previous section. Reading of Supporting Information Section S1 is encouraged for a more in-depth discussion on the linearity relation. In addition, some considerations about the limitations and the accuracy of the technique are debated in Supporting Information Section S3.

## 3. Results and discussion

### 3.1 OIS characterization of MIECs on YSZ for insertion-limited process

The aim of this section is validating the OIS by studying oxygen insertion in LSFx thin films on solid state electrolyte. LSFx thin films with two different Sr contents (x=0.5, 1) and an LSF/SFO bilayer were deposited on top of CGO-coated YSZ (001) substrates by Pulsed Laser deposition. For proper contact with the layers, a 10nmTi/100nmAu contact was around the LSF films. Silver (Ag) paste was applied on the backside of the samples as a counter-electrode (CE) for allowing oxygen exchange with the environment under the application of an electrochemical bias. A transparent capping layer of $Al_2O_3$ was deposited on top of the LSFx thin films to avoid oxygen insertion/evolution on thin films' surface (see Experimental section for the fabrication procedure). All the layers deposited present a single-phase structure with a smooth surface, ideal for the characterization of OIS (see Supporting information Section S4 for the microstructural and compositional characterization).

#### *3.1.1 Dynamic characterization of the state-of-charge in LSF50*

The exemplary case for OIS technique validation is the study of the chemical capacitance of LSF50 thin films. This material presents a well-known defect chemistry that allows us a direct evaluation of the capabilities of OIS.[25,39,40] OIS measurements were conducted at 350 ºC in air using the set-up described in Supporting information Section S4. The sample was first stabilized at different electrochemical potential between 0 and -0.425 V at 350 ºC, which corresponded to a range of oxygen partial pressures from 0.21 to $3.79\cdot10^{-15}$ bar following the Nernst potential Equation 11. No DC current was detected under these potentials, meaning that the capping layer effectively blocked oxygen insertion from the atmosphere through the LSF surface. At each of these biases ($E_1$), a sinusoidal voltage with an amplitude of $E_0$=±25 mV and frequency ranging from 0.01 Hz to 1 Hz was then applied to the LSF50 thin film. Evolution of optical properties of the sample was simultaneously characterized by ellipsometry after choosing the most suitable photon energy to be in the linear-approximation regime (see Supporting Information Section S1).

**Figure 3** shows a schematic cross section of the sample (**Figure 3a**) and experimental data of the OIS experiment. **Figure 3b and 3c** depict the Bode plot of the normalized module $|Y_{OIS,N}(\omega)|$ and the phase lag $\varphi(\omega)$ as a function of frequency derived from the parameters of the sinusoidal fitting (see Experimental section for more information about the normalization procedure). A general trend emerges for each of these two parameters: the relative amplitude of the optical oscillation, $|Y_{OIS,N}(\omega)|$, decreases with increasing frequency while the phase shift between the two signals $\varphi(\omega)$ increases with higher frequencies (note bigger errors at high frequency due to the synchronization error of the optical and electrical experiments). Moreover, the figure shows the relative optical response change and phase lag for measurements taken for four different voltage bias. Substantial differences can be observed in their response, a clear indicator of a change in the behavior of the film depending on the applied $E_1$. **Figure 3d** shows the Nyquist admittance plot of the optical admittance for the sample when measured at $E_1$=-0.075V. A semicircle is observed, closing at Re($Y_{OIS,N}(\omega)$)=1 and Im($Y_{OIS,N}(\omega)$)=0 for lower frequencies.

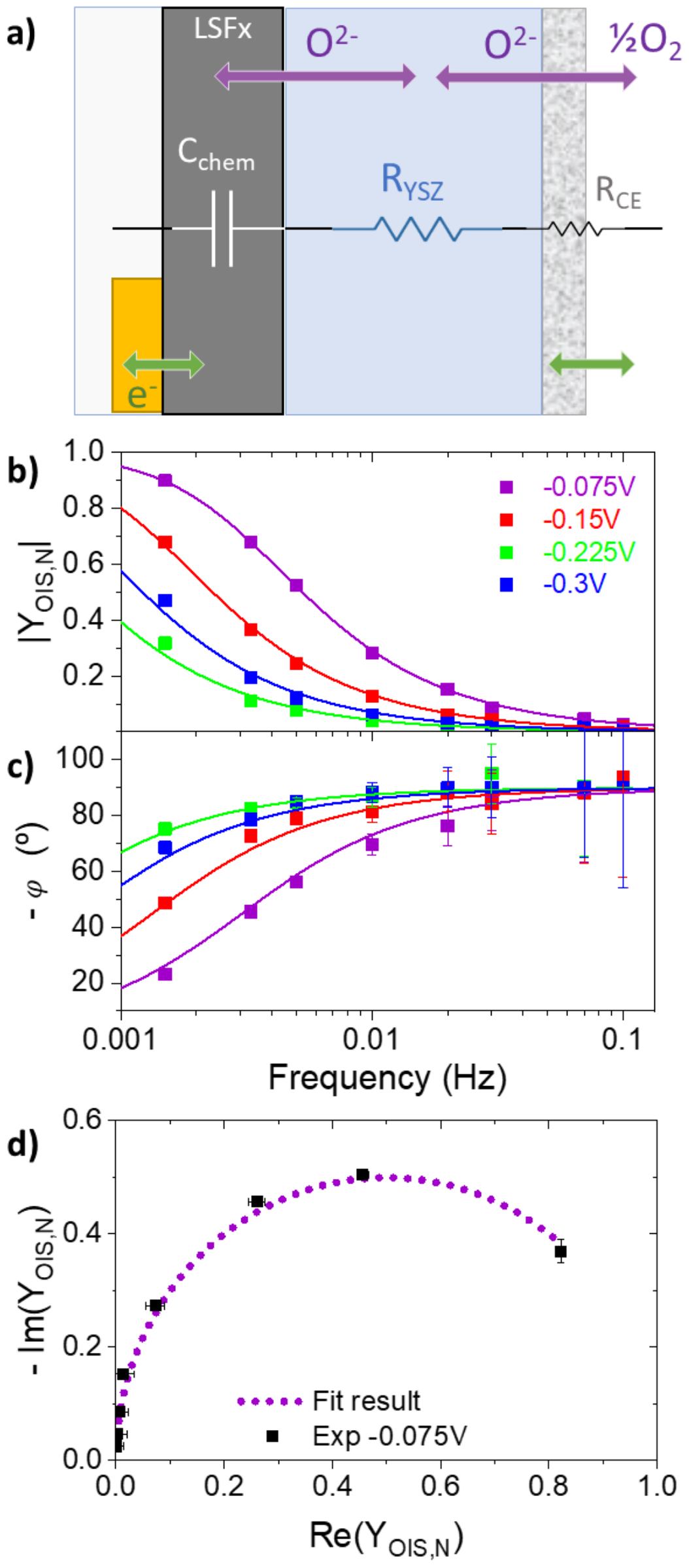

**Figure 3.** Bode plot presenting the change in the amplitude (b) and phase shift (c) as a function of frequency for four different $E_0$. d) Nyquist admittance plot of the $E_0$=-75 mV measurement and its fitting with the RC circuit model.

To properly fit the OIS spectra, it is necessary to consider the specific equivalent circuit of the electrochemical system under study in the case of both the EIS and OIS (see Supporting Information section S5) [38,41]. In the circuit, each component can be interpreted as an element. Starting from the Ag CE, a resistance $R_{CE}$ in parallel with a capacitance $C_{CE}$ is used to describe oxygen insertion on the counter-electrode and the double layer formation. The Alumina capping layer stops the oxygen insertion on the LSF surface; therefore, it can be considered as ion-blocking. Then, a resistance $R_{YSZ}$ in parallel with a capacitance $C_{CE}$ describes the oxygen conduction through the YSZ and the capacitance coming from the dielectric nature of the electrolyte, respectively. Finally, ambipolar oxygen motion inside layers is considered as the finite-space diffusion in 1D MIEC described in Section 2.2.2. Although the general form of the impedance is therefore quite complex (see Supporting Information section S2 for derivation), it can be drastically simplified in the present case. First, oxygen insertion and conduction in the CE and YSZ electrolyte generally present a fast time constant (*i.e.* τ=RC) and under the frequency range studied by OIS it can be approximated by a single resistive component, R. The circuit is therefore analogous to the one described in Section 2.2.2, with a resistance representing the oxygen insertion and motion through the YSZ/Ag in series with the diffusive distributed element. The similarity in the shape of Bode and Nyquist plot obtained in the OIS (Figure 3b,c,d) and the generalized OIS obtained in Section 2.2.2 for the case of low diffusion resistance (Figure 2f,g,h) suggests that the diffusion contribution is negligible compared to R. Indeed, simulations of the OIS circuit as a function of $R_i/R$ shows that, for the film's thickness and oxygen diffusivity constant expected at this temperature, the spectra do not show any significative contribution of the oxygen ion diffusion (see Supporting Information section S2 for the validation of this step).

Considering all this, the equivalent circuit of the response ends up an RC circuit (as the one presented in Figure 3a), where the chemical capacitance of the layer is charged by the insertion of ions from the YSZ and electrons from the current collector. Under these conditions one can obtain the EIS admittance as:

$$Y_{EIS}(\omega) = \frac{I(\omega)}{E(\omega)} = \frac{I(\omega)}{I(\omega)\cdot(\mathrm{Z_{LSFx}}+R)} = \frac{1}{\frac{1}{i\omega C}+R} = \frac{i\omega C}{1+i\omega RC} \quad (14)$$

In the case of OIS, where the observable parameter is the variation of optical properties in the LSF layer, we can take Equations 5, 7 and 8 and obtain an OIS admittance expression as:

$$Y_{OIS}(\omega) = \frac{\mathrm{Z_{LSFx}}}{\mathrm{Z_{LSFx}}+R}\cdot \mathrm{C}\cdot \mathrm{N} = \frac{\frac{CN}{i\omega C}}{\frac{1}{i\omega C}+R} = \frac{CN}{1+i\omega RC} \quad (15)$$

This formula gives the frequency response of the charge accumulated in the film for unit potential, i.e. the frequency dependent capacitive response of the system. As the analysis is performed with the normalized $Y_{OIS}(\omega)$, the result must be normalized. One can see that the normalized OIS admittance $Y_{OIS,N}(\omega)$ results in the transfer function of the capacitor element, i.e. the functional layer's transfer function:

$$Y_{OIS,N}(\omega) = \frac{1}{1+i\omega RC} \quad (16)$$

Using this relation, the impedance spectra were fitted, yielding a good representation of the experimental data for a single time constant RC, see Figure 3b,c,d. Moreover, since $R_{YSZ}>>R_{CE}$, the resistance R of the system can be approximated as the $R_{YSZ}$. This value can be calculated knowing the dimensions of the sample and the temperature. EIS measurements were used to confirm the validity of this approximation and the real value of $R_{YSZ}$ (see Supporting Information Section S5). This way, the chemical capacitance of the system can be obtained by OIS. This procedure was followed for all the measurements taken at voltages in the range from 0 to -0.425V and the obtained values for *Cchem* are shown in **Figure 4**. $C_{chem}$ changes depending on the applied voltage and a peak around 6500 F/cm$^3$ can be observed at an applied bias of -0.225 V which corresponds to a $pO_{2,eq}$ on the order of $1.1\cdot10^{-8}$ bar. Importantly, a good agreement is observed between $C_{chem}$ extracted by EIS and OIS, validating the procedure (see Supporting Information Section S5).

The calculated values of *Cchem* can be used to determine parameters of the defect chemistry of the material under study. According to the chemical capacitance model for LSF with dilute defects reported by Schmid *et al.*, [42,43] the relation of point defects and *Cchem* is given by:

$$C_{chem_{LSF}} = \frac{e^2}{kT} n_{uc} \left(\frac{1}{4c_v} + \frac{1}{c_h}\right)^{-1} \quad (17)$$

Where $c_v$ and $c_h$ denote the concentration of oxygen vacancies and electron holes in the LSF50 thin film and $n_{uc}$ is the number of unit cells. Note that this equation can only be used considering both the range of equivalent oxygen pressure and the temperature conditions used in this work, as the concentration of $Fe^{2+}$ electrons is expected to be negligible.[39,40,42,43] The red line in **Figure 4** shows the result of the fitting of the equilibrium constant $K_{ox}$ for the oxygen insertion reaction of Equation 12 when considering both Equation 11 and 17. The final obtained value is $K_{ox}=5\cdot10^4$, which is in the order of reference values. [25,39,40] For the obtained value of Kox the evolution of $[O_O^x]$ with the applied bias can be extracted from Equation 17 (see blue curve in Figure 4).

In summary, the validation of OIS was achieved by assessing the response of the LSF50 film to various voltage inputs. Both Optical and Electrical Impedance Spectroscopy $C_{chem}$ values exhibit the same pattern, and the technique has allowed for the extraction of the defect chemistry of LSF50. An analysis of the same sample employing a device-oriented approach has also been done in Supporting Information Section S6, including a response time and performance point of view analyses.

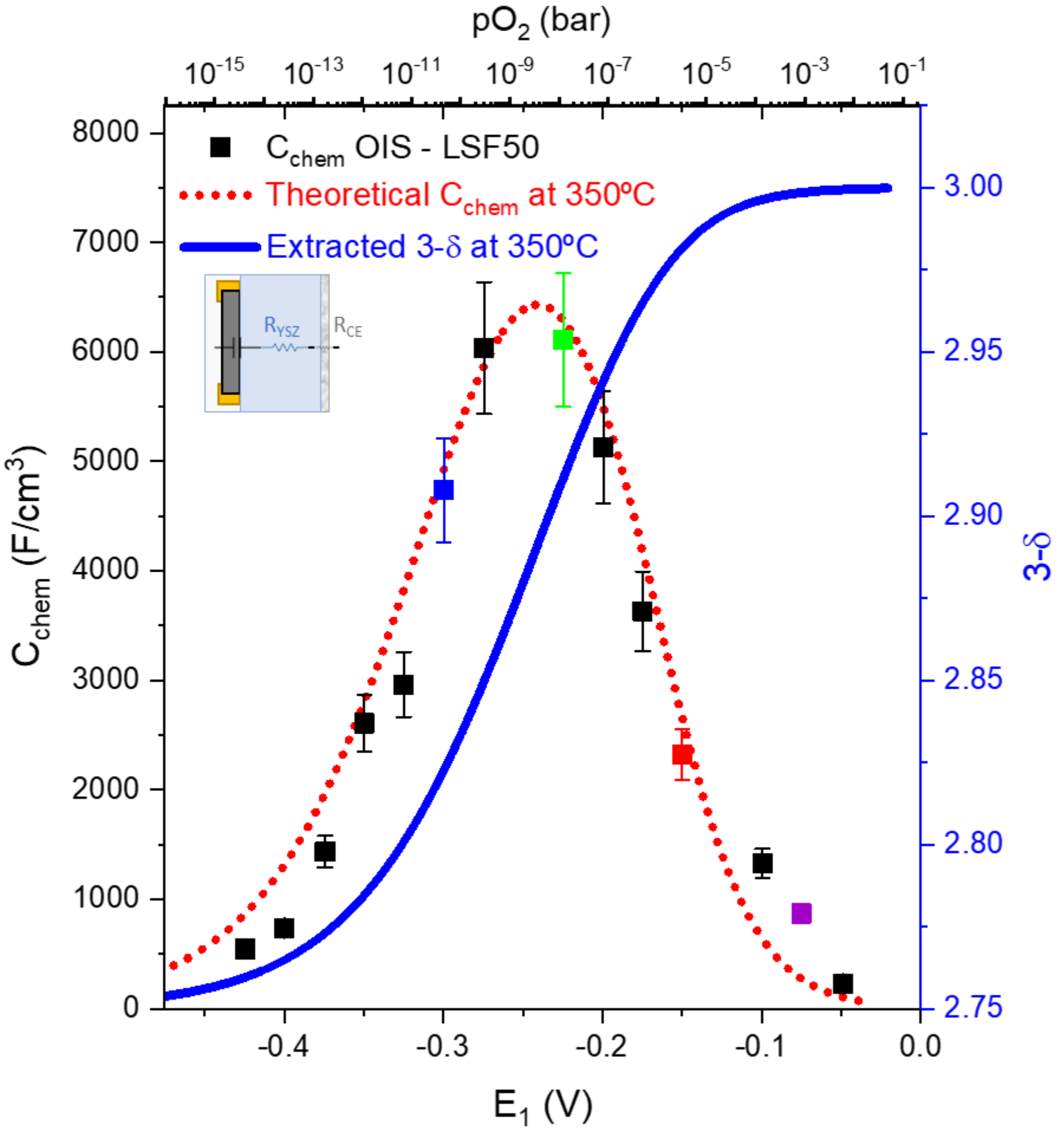

**Figure 4.** Chemical capacitance measured via OIS for the different applied $E_1$. The fitting of the defect chemistry model is shown in red. The graph also shows the resulting relation between V (and $pO_2$) with the oxygen content per unit cell in blue.

*3.1.2 In-situ characterization of a topotactic transition material by OIS*

After validating the technique in the previous section with a material with a known defect chemistry as LSF50, we moved the analysis to the characterization of SFO thin films, a material with a complex defect chemistry. At the experiment's operating temperature of 350°C, SFO undergoes a topotactical phase transition so that its structure is determined by the oxygen content, *i.e.* the electrochemical potential, resulting in either perovskite (PV), brownmillerite (BM), or a combination of the two phases.[44] To demonstrate the capabilities of OIS in studying this interesting phase transition, a 60 nm-thick SFO film was studied (see Experimental section for the fabrication procedure as well as Supporting Information Section S4 for the microstructural and compositional characterization). OIS experiments were carried out following a similar procedure as in the LSF50 sample. In this case, as the number of vacancies that the layer can accommodate is significantly increased the lowest frequency used was 1 mHz. Again, linearity of the signal at low enough applied biases was confirmed prior to the experiment and the sample was stabilized at voltages in the same range of the previous section before each OIS spectrum was acquired.

**Figure 5a** and **5b** show the Bode plots obtained for the SFO layer at different $E_1$. In line with the findings for LSF50 shown in Figure 3b and 3c, at higher frequencies the relative amplitude $|Y_{OIS,N}|$ decreases while the phase shift $\varphi(\omega)$ increases, following the expected behavior for the RC circuit model implemented in the previous section. Moreover, the Nyquist admittance plot for the measurement at -0.075V and illustrates the semicircle, which is gradually converging toward Re($|Y_{OIS,N}|$)=1 (see **Figure 5c**). All the simplifications adopted for the LSF50 to the general model are still valid for SFO, which reduces $Z_{SFO}$ to a capacitive element. Nevertheless, the direct comparison between the Bode plots of LSF50 (Figure 3b and 3c) and SFO (Figure 5a and 5b) reveals interesting differences in the behavior as a function of $E_1$. These results motivate the fitting on the signal in order to calculate the $C_{chem}$ of the material as a function of the applied bias or the equivalent $pO_2$. These values are reported in **Figure 5d**. The material's $C_{chem}$ is constant for more oxidizing voltages (near 0 V) and has a sudden increase around $E_1$= -0.185V before decreasing again for more reducing voltages. These results are consistent with a sharp peak of chemical capacitance expected in the case of an oxygen-induced phase transition.[45] Indeed, literature reveals a distinct three-regime relationship between oxygen vacancies and $pO_2$: firstly, at low $pO_2$ levels, the material adopts the Brownmillerite (BM) structure, stabilizing its stoichiometry due to the organized distribution of vacancies within the unit cells; secondly, a sudden surge in oxygen content within the structure (resulting from vacancy, $V_O^{\cdot\cdot}$, reduction) occurs as BM ceases to be the favored structure, signaling the onset of a phase transition to PV; finally, when the material is in the PV phase, there is a gradual increase in $[O_O^x]$.[44,46,47]

Since a dilute model cannot describe such complicated defect chemistry, the variation of the oxygen concentration in SFO can be calculated from $C_{chem}$ considering the definition of this parameter shown in Equation 17. First, experimental values of $C_{chem}$ are fitted by two exponential functions to describe the rapid upsurge in the central region. The final fitted curve following this method is shown in Figure 5d. Then, by integrating the $C_{chem}$ relation with voltage, the oxygen concentration of the material can be obtained (plotted in blue in Figure 5d). Here, we considered that the oxygen concentration of BM-SFO for the fully reduced state is 2.5 (as no $Fe^{2+}$ electrons are considered in these conditions). As anticipated by the existing literature on SFO behavior, the sudden alteration in the quantity of oxygen vacancies when the phase transition takes place produces a sharp peak in $C_{chem}$. At the studied temperature, this transition happened at a voltage of -0.185V, which corresponds to an equivalent partial oxygen pressure of $pO_2=2.23\times10^{-7}$ bar, consistent with the trend observed in literature.[47] Also, the oxidation curve of the material is very much in line with the expected behavior observed in literature.[44,46,47] Before the transition to the PV, oxidation of the material is difficult, as the BM phase is better stabilized when δ is near 0.5. When the topotactic transition is complete, the oxidation occurs in a gradual way, almost proportional to the applied voltage i.e., exponentially increasing with $pO_{2,eq}$ according to the Nerst potential Equation 11. This phenomenon can potentially be described by a non-dilute model as some references propose.[48] Finally, the δ at which this transition starts to take place when reducing the material is δ~0.3 (i.e., [Ox] ~2.7) falls in the stabilization range proposed by Mizusaki et al.[44]

As a conclusion, we demonstrate that OIS technique can be used to obtain the concentration of oxygen vs potential in a material with a complex defect chemistry such as SFO. This validates that the presented method is a powerful technique for the in-situ characterization of the oxidation curve of materials that present a change in optical properties. Unlike other techniques, this has been done by employing a modeless approach, as no optical model is needed during the analysis, obtaining results that agree with expectations.[44–47,49]

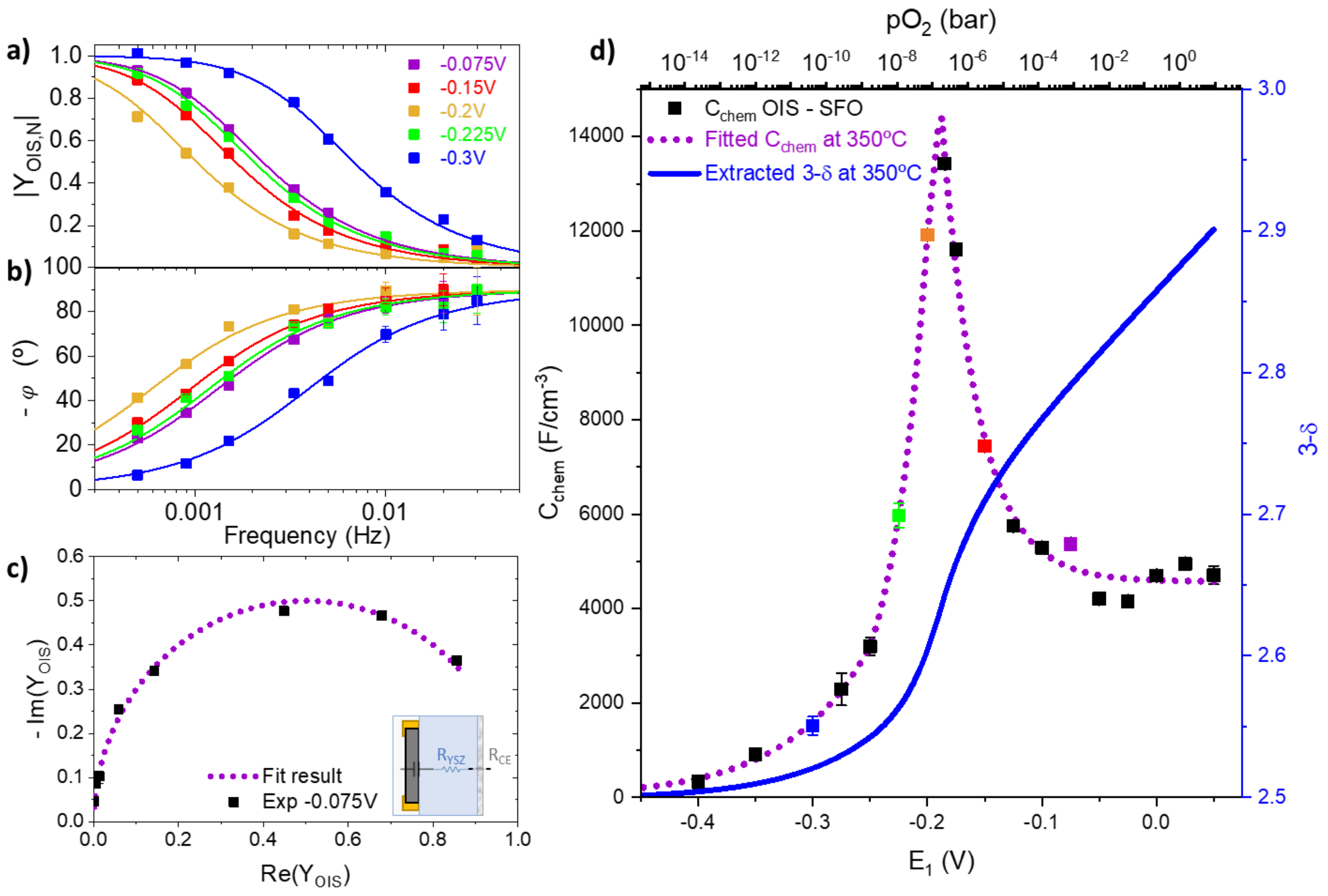

**Figure 5.** a-b) Bode plots for different applied $E_1$ in the SFO sample. c) Nyquist plot for the measurement performed at $E_1$=-0.075V with its corresponding fitting with the RC model. d) Experimental Chemical capacitance of the material calculated with OIS. Dashed purple line shows the fitting of the capacity vs voltage and right axis in blue shows the oxidation state depending on the applied voltage (and equivalent $pO_2$).

*3.1.3 In-situ charging of a multilayer of materials with dissimilar dynamic behavior*

Real application systems do not often consist of monolayers but in more complex architectures. With the aim to show the possibilities of OIS, a bilayer battery-like device was studied. As the optical properties of multilayer systems are traditionally complicated to retrieve by spectroscopy ellipsometry, we aim at demonstrating the advantages of using a modeless technique such as OIS to quantify the variation of ion concentration. The structure of the employed sample was conserved but this time a layer of SFO was deposited on top of the LSF50 layer prior to the capping with $Al_2O_3$. The LSF50 and SFO layers have thicknesses of 18 nm and 22 nm, respectively. OIS experiments were carried out following the same procedure used in the previous sections. The measurements were carried out with the light beam on top of both layers to measure the change of both materials simultaneously.

In order to get more insights into the contribution of the two layers to the measured capacitance, it is useful to recognize that the overall impedance can be represented as the series of two distributed element according to the model derived by Jamnik and Maier, [38,41] as presented in section 3.1.1 (see **Figure 6a**). Since no diffusion limitations are expected on both layers under the conditions studied, the ionic resistances shown in the equivalent circuit of Figure 6a can be neglected and the bilayer is then described as two capacities in parallel one for each layer, which charge simultaneously (see **Figure 6b**). Therefore, the final circuit becomes straightforward as these two capacitances can be treated as a single capacitance obtained by summing them together. This entire analysis and the simplifications made are essential because they enable us to interpret the system using the same circuit that we employed in the previous sections. As the capacity of each of the layers $C_{LSF50}$ and $C_{SFO}$ is obtained by multiplying the chemical capacitance by the volume of each film, the total capacitance is equal to the weighted sum of the chemical capacitances:

$$C_T = C_{LSF50} + C_{SFO} = v_{LSF50} \cdot C_{chem_{LSF50}} + v_{SFO} \cdot C_{chem_{SFO}} \tag{18}$$

Where $C_T$ is the total capacity of the bilayer, $v_{LSF50}$ and $v_{SFO}$ are the thin film corresponding volumes and their respective volumetric chemical capacitance is noted as $C_{chem_{LSF50}}$ and $C_{chem_{SFO}}$. By applying the same procedure of LSF50 and SFO, an overall capacitance of the bilayer can be extracted as a function of the applied bias $E_1$. **Figure 6c** shows the experimental $C_T$ values obtained via OIS. The data shows a clear peak at around $E_1$= -0.185V consistent with the $C_{chem}$ peak of SFO (see Section 3.1.2), and a shoulder is observed at $E_1 \approx$ -0.25V also in agreement with the findings on $C_{chem}$ for LSF50 (see Section 3.1.1). The values obtained using OIS are also in good agreement with the ones extracted from the EIS analysis. The figure also presents the shape of the expected $C_{LSF50}$ and $C_{SFO}$ from the data shown in Figure 4 and Figure 5d as well as the calculated $C_T$ as the sum of the precedent. One can observe a good with experimental $C_T$, especially when measuring on the extremes of the experimental voltage range used. When approaching more negative (reducing) voltage values, $C_{LSF50}$ describes well the measured capacity of the sample. When approaching more positive (oxidizing) voltages, $C_{SFO}$ dominates the overall bilayer behavior, as expected from Equation 18. Nevertheless, for intermediate voltages where both capacities are similar, the experimentally measured $C_T$ clearly follows the trend expected from the sum of both contributions. Minor differences are seen between experimental and theoretical behavior of the OIS signal. It is known that reducing the thickness of the films, materials are more susceptible to interface effects.[50,51] Despite the functional films' thickness difference with respect to the previously studied samples, this did not seem to be the case, as no important change in the oxidation curves of the materials was observed. These results demonstrate the effective use of OIS for analyzing composite elements as the theoretical response of the system to an external voltage stimulus was consistent with experimental observations.

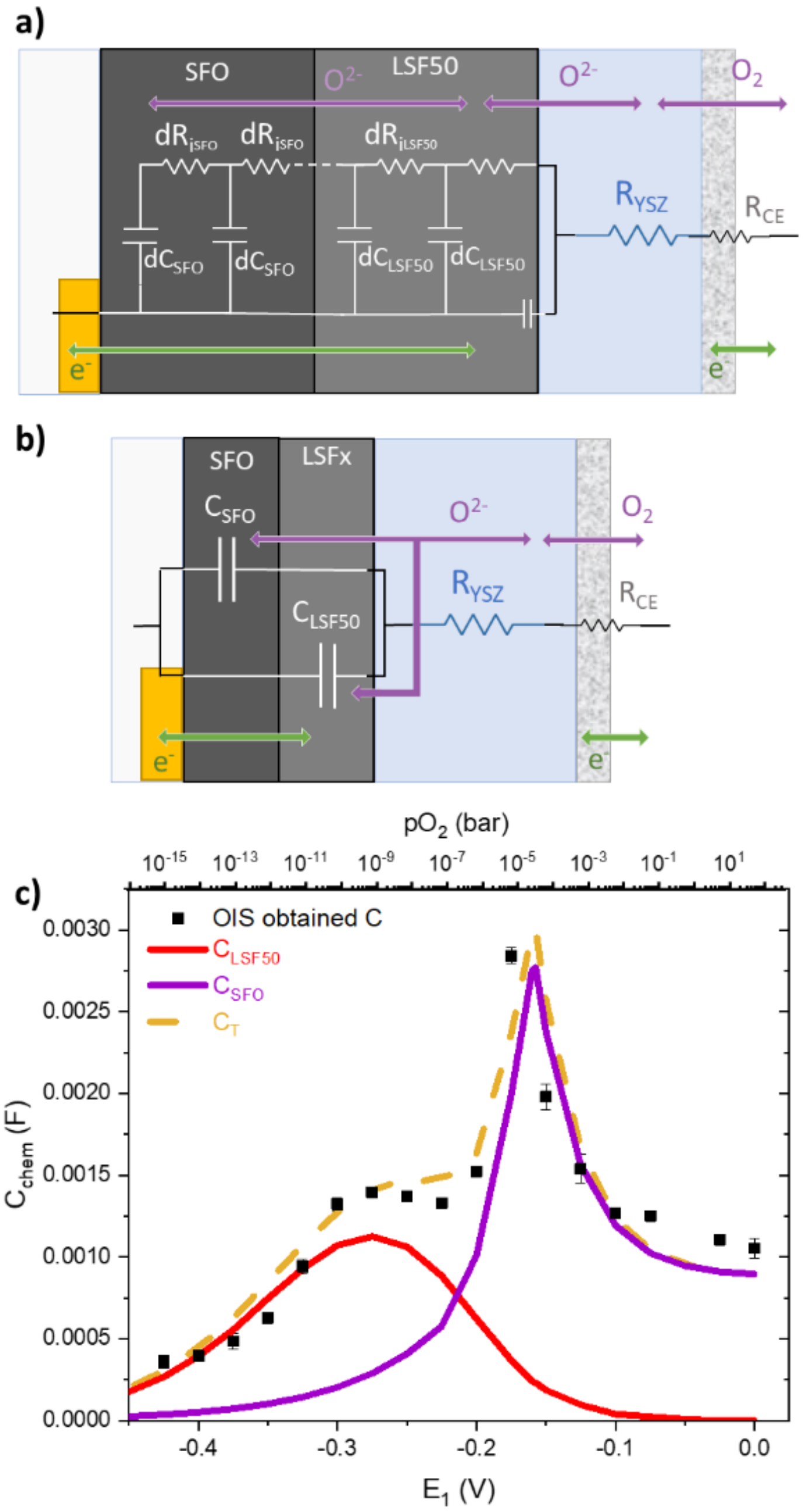

**Figure 6.** a) and b) show the equivalent circuit for the EIS and OIS analysis for a diffusion limited case and a surface limited charging for the bi-layer system, respectively. c) Experimental chemical capacitance calculated with OIS. The purple line shows the fitting of the expected $C_{SFO}$ and red line the one for $C_{LSF50}$.

### 3.2 Diffusion-limited OIS characterization of LSF50 in a liquid electrolyte

To fully demonstrate the potential of the technique, OIS was employed to characterize ion diffusivity. Real systems working with ionic motion, such as batteries or solid oxide fuel cells, have performance limitations due to ion diffusivity. Studying the ionic charging limitations due to low mobility of charged species is of utmost importance for improving device's capabilities. For this reason, we decided to investigate ion diffusivity of LSF50 thin films measured at room temperature, using a 0.1 M KOH solution as liquid electrolyte. We previously showed that the defect concentration in LSF50 thin film can be varied at RT by the application of an electrochemical potential in alkaline electrolytes, through a mechanism that can be described by a dilute defect chemistry model.[26] For this purpose, a 120 nm thick LSF50 thin film was deposited on top of a platinum (Pt) coated silicon chip. A home-made electrochemical chamber is employed for the characterization of the sample in a liquid-electrolyte electrochemical cell. The chamber is equipped with two transparent optical windows that are perpendicularly tilted to the incident light beam in order to ensure full optical transmission of the light beam. In-situ SE measurements can be carried out using a three-electrode configuration (see **Figure 7a**), in which a counter electrode (Pt mesh), a reference electrode (Ag/AgCl) and a liquid electrolyte are introduced into the chamber while the working electrode is contacting the Pt of the sample (see schematic in **Figure 7b**). Oxygen insertion reaction is favored by the use of a 0.1M KOH alkaline electrolyte solution.

Figure S8 present the cyclic voltammetry of the sample measured at a cycling speed of 0.5 mV/s. The redox reaction peaks centered at $E_{1/2}$ = -0.2 V applied versus the Ag/AgCl reference are observed in the CV curve for the LSF50 thin film, which is consistent with previous results.[26] Interestingly, the CV scans present a certain asymmetry that are likely originated by kinetic limitations, and precisely linked to the low ionic diffusivity in the material at room temperature. Similar to the OIS experiment presented for the previous samples, linearity of the signal at low enough applied biases was confirmed and experiment was run for a single $E_1$= -0.2 V. **Figure 7d and e** show the resulting Bode-$\varphi(\omega)$ and Nyquist plots for the OIS admittance. As for the insertion limited cases, there is a decrease in the relative amplitude $\Psi_N(\omega)$ and a phase shift $\varphi(\omega)$ increase when moving to higher frequencies. Nevertheless, a distinct behavior is seen, as the phase shift $\varphi(\omega)$ does not directly increase up to 90º but seems to saturate in intermediate frequencies at values near 45º. This feature is similar to the expected behavior for the diffusion-limited 1D model introduced in Section 2.2.2 (see Figure 2).

To understand this behavior prior fitting of the OIS spectra, EIS measurements performed at the same $E_1$ were subjected into analysis as well. When analyzing the equivalent electrical circuit, the increase of complexity is apparent. Usually, ionic insertion into materials through a liquid electrolyte is modelled by the Randles circuit (**Figure 7c**). This circuit includes three different processes occurring in the sample. At first, a series resistance deriving from the electrolyte, $R_{ely}$. Then, oxygen insertion and diffusion into the material are modelled by the parallel combination of the double-layer capacitance, $C_{dl}$, and the insertion resistance, $R_{inc}$, followed by the previously mentioned Warburg element ($Z_W$) describing the film. This model can be used to describe the observed EIS spectra shown in Supporting Information Section S7. Nevertheless, as $R_{ely} \ll R_{inc}$, this initial term can be neglected, and the circuit directly can be simplified as the one shown in Figure 2a for the diffusion limited case. This approximation can also be done in frequency terms, as the response time of the double layer charging $\tau_{dl} = R_{inc}C_{dl}$ is orders of magnitude lower than the one for the layer charging $\tau_{layer} = (R_{inc} + R_i)C$, when analyzing the circuit at lower frequencies, it can be reduced to the long time response line.

Figure 7d and 7e show different theoretical calculations of normalized equation 10 showing the effect of a decreasing ionic diffusion represented by different $R_i$ values for a given $R = R_{inc}$=1 kΩ and $C$= 0.02-0.04 F. Simulations clearly show the effect of the diffusion in the layer as they are able to reproduce the shape of experimental data. The relation between ionic resistance of the film and its ionic diffusivity, $D_o^q$, is given by:

$$D_o^q = \frac{t}{R_i \cdot A} \cdot \frac{k_b T}{c_{O2} \cdot (z_i \cdot e)^2} \qquad (19)$$

Where L and A are respectively the thickness and the surface area of the LSF and $c_{O2}$ is the concentration of oxygen in the lattice. The diffusivity values follow the experimental trend are in the range of $1.86 \cdot 10^{-15}$-$6.26 \cdot 10^{-16}$ $cm^2 s^{-1}$. These diffusion coefficients match with previously reported values for LSF films.[52]

Despite the potential of OIS to clearly see the effect of diffusion limitation in the layer, small differences between theoretical and experimental data are seen. Usually, non-ideal Warburg elements as well as constant phase elements (CPEs) are used to account for non-idealities in EIS. The origin of these non-linearities can be as varied as different diffusion lengths, surface heterogeneity, or secondary reactions with the working electrode. A potential explanation of the discrepancy between experimental data and ideal behavior is the presence of multiple species insertion. Previous work has shown that not only oxygen can be inserted into LSFx via liquid gating in KOH electrolyte, but also $H^+$.[26]

Despite these non-idealities, OIS could be used in the study to measure the response of a material when being ionically (dis)charged via voltage application and to describe the behavior of the system at low frequencies. As a matter of fact, some benefits of OIS precisely reside on this last point. EIS measurements are usually affected by substantial error at low frequency, as the current minimizes. Meanwhile, OIS becomes more accurate for the low frequency range as the capacitance of the film gives its the maximum response. Moreover, EIS measures the total current through the system and so, it may present parallel contributions from other capacitive elements, such as reactions on the Pt current collector, while OIS technique is only sensitive to the change in the ions (de)inserted in the optically observed material.

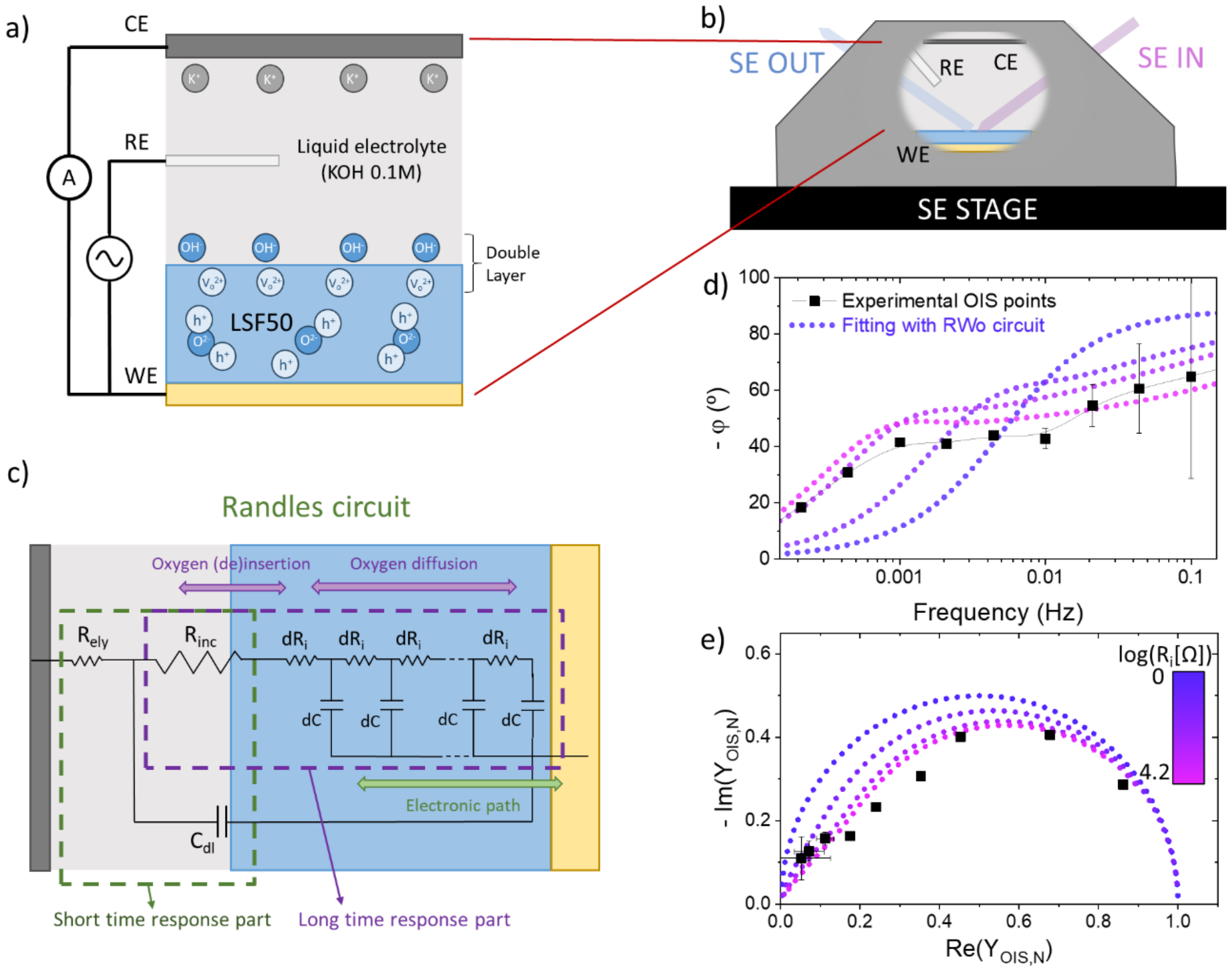

**Figure 7. a)** Schematic of the system for the three-electrode liquid electrolyte gating. **b)** Set-up schematic showing the electrochemical cell on top of the Spectroscopic Ellipsometry stage, light entering ang exiting the chamber after reflecting on the sample, and the different components of the system. **c)** Schematic of the Randles Circuit to simulate the oxygen ion insertion from the electrolyte. **d)** Bode Admittance and **e)** Nyquist Admittance plots for the OIS experimental data and different theoretical calculations of the OIS behavior for different oxygen diffusion values.

## 4. Conclusion

Optoionic Impedance Spectroscopy (OIS) has been presented in this work as a novel operando technique for the characterization of ionic insertion in functional thin films. First, we developed a theoretical framework to analyze the OIS response of MIEC materials, highlighting that OIS directly measures the variation of chemical charge upon the application of an electrochemical potential. Then, we demonstrated the capability of OIS in studying the oxygen ion variation in LSF50 and SFO thin films on solid state electrolytes. In these systems, OIS is insertion limited and it can be used to quantify the chemical capacitance of the layers as a function of the electrochemical potential applied. Moreover, OIS has shown its capability to effectively measure complex layers composed by more than one material, overcoming the limitations of optical fitting of traditional SE. The application of the technique can be further expanded for studying device response characteristics to gain deeper insight of the mechanisms involving the change and how to optimize the behavior to controllably change its functionalities. More complex systems involving diffusion-limited charging were also analyzed using OIS, through cycling LSF50 thin films at RT in liquid electrolytes. The results proved that the theoretical approach of the OIS can be employed to extract relevant parameters as ionic motion inside functional oxide films. The advantages of OIS are the modeless analysis of the results that make it applicable for the study of a wide variety of samples as long as any optical feature is modified by ion insertion.

**5. Experimental Section**

*Sample fabrication:* $La_{1-x}Sr_xFeO_3$ (LSFx) thin films of two different Sr concentrations (LSF50 and SFO with 50% and 100%, respectively) were fabricated using Pulsed Laser Deposition (PLD) on gadolinium-doped ceria (CGO)-coated YSZ (001) 1x1 $cm^2$ substrates with a thickness of 0.5 mm. The CGO layer (approx. 20 nm thick) was deposited on the YSZ substrates a barrier layer, preventing the formation of secondary phases at the LSF/YSZ interface.[53,54] For the diffusion-limited study, an LSF50 film was deposited following the same methodology on a 400nm $SiO_2$ / 10nm Ti / 70nm Pt coated Si 1x1 $cm^2$ chip. All the layers were deposited employing a large-area system from PVD products (PLD-5000) equipped with a KrF-248 nm excimer laser from Lambda Physik (COMPex PRO 205). A commercial pellet of CGO was used as target material, while the LSF50 and SFO were prepared by solid state synthesis following the process explained in reference [25]. During all the depositions, the substrate-target distance was set to 90 mm and the substrate was kept at 600 °C in an oxygen partial pressure of 0.0067 mbar. The laser frequency employed was 10 Hz and the energy fluency was 0.8 J $cm^{-2}$ per pulse. The first sample consists of ~100 nm of LSF50, the second one of ~60 nm of SFO as measured by ellipsometry and Atomic Force Microscopy (AFM). A third sample was also produced containing a double layer of ~18 nm of LSF50 and ~22 nm of SFO**.** The deposition of these functional layers was done using a microfabricated Si hard mask with a size of 2.5x3.5 $mm^2$ centered in the substrate. To electrically contact the functional thin films, a conductive 10nm Ti/100nm Au electrode mask was microfabricated by photolithography and thermal evaporation. The thin Ti layer in the contact is deposited in order to promote a good adhesion of the Au to the sample. To prevent the equilibrium of the functional layer with the atmosphere when going to higher temperatures, a last capping layer of alumina ($Al_2O_3$) of approximately 100 nm was also deposited via PLD (see a schematic of the samples in Supporting Information Section S4). Alumina was chosen as the capping material because of its low oxygen mobility and negligible oxygen stoichiometry variation in a wide range of oxygen partial pressure ($pO_2$) and temperatures.[55] Another important property of the alumina capping is its low absorption, which allows the measurement of the optical properties of the functional layer with more precision. Silver paste was used on the backside of the YSZ substrate as oxygen counter electrode. For the second case of study, the LSF50 deposited on top of the Pt coated Si chip had a thickness of ~120 nm.

*Structural characterization:* Structural analysis of the thin films was based on high-resolution X-rays Diffraction (XRD) measurements carried out in a coupled Θ–2Θ Bragg–Brentano configuration using a *Bruker D8 Advanced diffractometer* equipped with a Cu Kα radiation source. The topography of the LSF thin films as well as the deposition thickness was characterized in non-contact mode by Atomic Force Microscopy (AFM) using the *XE 100 model of Park System Corp*. Cross-section image taken after sample measurement was taken using a *Zeiss Auriga* Scanning Electron Microscopy (SEM).

*OIS measurement experimental and considerations:* In this paper, operando ellipsometry measurements with *Horiba UVISEL Plus Spectroscopic Ellipsometer* for the solid-electrolyte battery-like cells were performed on the samples in air at 350C under real electrochemical conditions. The experimental setup for in-operando ellipsometry measurements, illustrated in the schematic of Supporting Information Section S4, includes a Linkam stage (*THMS600*) as the substrate holder that enables the temperature control and electrical connections, as well as a cover made by aluminum film that allows the light beam to pass through while reducing heat dissipation. The polarized light beam enters the stage at an angle of incidence of 70º and is focused on the sample with a spot size of 2 mm$^2$. After the interactions with the sample, the light beam is reflected and read by the detector, and the optical spectra are obtained in a photon energy range from 0.6 eV to 5.0 eV using monochromatic ellipsometry. DC voltage bias ($\Delta V$) and AC voltages were applied between the LSF layer (through the gold contact working electrode) and the silver counter electrode using a potentiostat of the model of *SP-150 from Biologic* in order to tailor the equivalent oxygen partial pressure in the LSF thin films ($pO_{2,eq}$) according to the Nernst potential. The chosen voltage window of -0.45V to 0V was chosen deliberately to be around the oxygen insertion potential at the chosen experiment temperature. In the case of the diffusion-limited study, a home-made chamber fabricated by 3D printing using a *Prusa i3 style 3D printer* and *Acrilonitril Butadiene Styrene* was employed for the characterization of liquid-electrolyte electrochemical cells. The chamber is equipped with two transparent optical windows which are perpendicularly tilted to the incident light beam in order to ensure full optical transmission of the light beam. Besides, the chamber architecture is adapted to the introduction of the reference electrode. The LSF50 thin film was attached to a copper wire using graphite paste and its edges were covered by a robust epoxy resin glue from UHU to avoid undesirable reactions, leaving an uncovered LSF50 area of 72 $mm^2$. The electrochemical measurements were carried out on a three-electrode configuration in which: the sample was the working electrode (WE) an aqueous 0.1 M KOH solution was employed as an electrolyte and a Pt mesh was introduced as a counter electrode (CE). The reference was an Ag/AgCl commercial reference. Cyclic Voltammetry (CV) and OIS experiment were performed at room temperature using the same *SP-150 Biologic* potentiostat.

Before each measurement, the sample was stabilized applying a DC voltage step at the maximum and minimum voltages it will go through during the experiment ($E_1 \pm E_0$) and waiting for equilibrium. At each of this voltages, Ψ values were noted as $\Psi_1 + \Psi_0$ and $\Psi_1 - \Psi_0$ for the maximum and minimum applied voltages, respectively. Linearity between the measured $\Psi_1 + \Psi_0$, $\Psi_1$ and $\Psi_1 - \Psi_0$ with their respective voltage was checked before initiating the OIS measure. Once this was done, the value of the experimental optical admittance was further normalized by dividing the electrical and optical signals by their maximum amplitude value. In the case of the optical signal this was done by first calculating the maximum change of Ψ:

$$|\Delta\Psi|_{max} = [\Psi_1 + \Psi_0] - [\Psi_1 - \Psi_0] = 2\Psi_0 \quad (20)$$

As the electrical signal measured has a constant amplitude during the experiment, the data for normalized admittance analysis came from the relative decrease/increase of the optical signal as:

$$Y_{OIS,N}(\omega) = \Psi_N(\omega)\big(\sin(\varphi(\omega)) + j\cos(\varphi(\omega))\big), \;\; where\; \Psi_N(\omega) = \frac{\Psi(\omega)}{\Psi_0} \quad (21)$$

Where $Y_{OIS,N}(\omega)$ and $\Psi_N(\omega)$ are the normalized admittance and the relative amplitude of the optical signal.

**Supporting Information**

Supporting Information is available from the Wiley Online Library or from the author.

**Acknowledgements**

This project received funding from the European Union's Horizon 2020 research and innovation program under grant agreement No. 824072 (HARVESTORE), No 101066321 (TRANSIONICS) and, No. 101017709 (EPISTORE) and under the Marie Skłodowska-Curie Actions Postdoctoral Fellowship grant (101107093). The authors acknowledge support from the Generalitat de Catalunya (2021-SGR-00750, NANOEN).

Received: ((will be filled in by the editorial staff))

Revised: ((will be filled in by the editorial staff))

Published online: ((will be filled in by the editorial staff))

# Supporting Information

**Optoionic Impedance Spectroscopy (OIS): an in-situ modeless technique for electrochemical characterization of mixed ionic electronic conductors**

*Paul Nizet, Francesco Chiabrera*, Yunqing Tang, Nerea Alayo, Beatrice Laurenti, Federico Baiutti, Alex Morata, Albert Tarancón**

**S1: Importance of signal linearity for Optoionic Impedance Spectroscopy analysis**

The OIS outlined in this paper relies on establishing a linear relationship between applied voltage and the measured property under equilibrium conditions. Throughout the analyses conducted on the various samples, the choice of ellipsometer output (Ψ) was intentionally made to maximize the linearity of the relationship between these two signals. **Figure S1a** illustrates that expecting complete linearity across all photon energies or voltage ranges is not attainable. This observation aligns with the multitude of factors influencing the output value.[1] Each of the ellipsometry dispersion parameters is contingent upon the interactions between light and the different materials i.e., with their refractive index (n) and extinction coefficient (k), along with light interference arising from reflection at the surface and each interface. Having multi-layer samples includes more complexity in the change of the measured parameters, making the linear change relation quite difficult to happen. Moreover, it's important to highlight that the change in oxidation state does not need to exhibit linearity with voltage. Within the materials being studied, this relationship is in fact non-linear, introducing an additional level of complexity.

**Figure S1b** depicts OIS measurements simultaneously acquired at various photon energies ($h\omega$ = 2.564 eV, 2.956 eV, and 3.285 eV). To achieve a linear relationship between signal and voltage, it is essential that the maximum ($E_0+E_1$), minimum ($E_0-E_1$), and intermediate ($E_0$) voltages exhibit linear correspondence with the ellipsometer signal $\Psi(h\omega)$. The cases presented in the figure, ordered by increasing energy, exhibit three scenarios: a linear relationship, a relationship close to linearity, and a non-linearity relation. In the first scenario, where $\Psi(2.564eV)$ values for the three specified voltages are evenly spaced, the resulting output signal forms a sinusoidal waveform. In the second case, the $\Psi(2.956eV)$ value increases with voltage but not linearly; in other words, the Ψ values for each of the three mentioned voltages are not equidistant. The measured signal in this case displays an asymmetric sine wave. In the final scenario, a distinctly different relationship is observed. Here, the $\Psi(3.285eV)$ value at equilibrium with the middle voltage is lower than the Ψ values at the maximum and minimum voltages. Consequently, a function is noted where Ψ increases as it approaches either of the extreme values within the experiment.

As elucidated in the paper, the measurement analysis is conducted by interpreting both the input and output as sinusoidal signals. This simple process enables the derivation of the chemical capacity, $C_{Chem}$, value using the corresponding circuit model. **Figure S1** illustrates how, with the appropriate photon energy, we are positioned within area A (depicted in green) after the sinusoidal adjustment where fitting can be performed on the relative amplitude and phase shift. However, given that the technique relies on examining these two variables as functions of frequency, it is also viable to manually measure the amplitude of oscillation $\Psi_0(\omega)$ and the delay time in the optical response to extract the phase shift $(\omega)$. This manual approach is sufficient for conducting the analysis. Thus, it becomes evident that, despite not exhibiting a strictly linear relationship, the analysis remains valid as long as these two parameters can be manually extracted. In this manner, for the second scenario, even if analyzing the signals as sine waves through fitting is not feasible, OIS analysis can still be carried out by manually extracting those two parameters. The outcome significantly differs for the third scenario. In this particular case, it is not possible to accurately determine the amplitude change of the oscillation, rendering the analysis unfeasible. While it is feasible to estimate the delay time by using the peaks of the Ψ signal as a reference, this approach becomes highly challenging at higher frequencies due to the increased prominence of signal non-linearity. Nonetheless, employing OIS fitting through phase and amplitude offers improved precision in $C_{Chem}$ extraction. Consequently, an analysis relying solely on phase shift may be sufficient for its calculation but may fall short in terms of achieving minimal error.

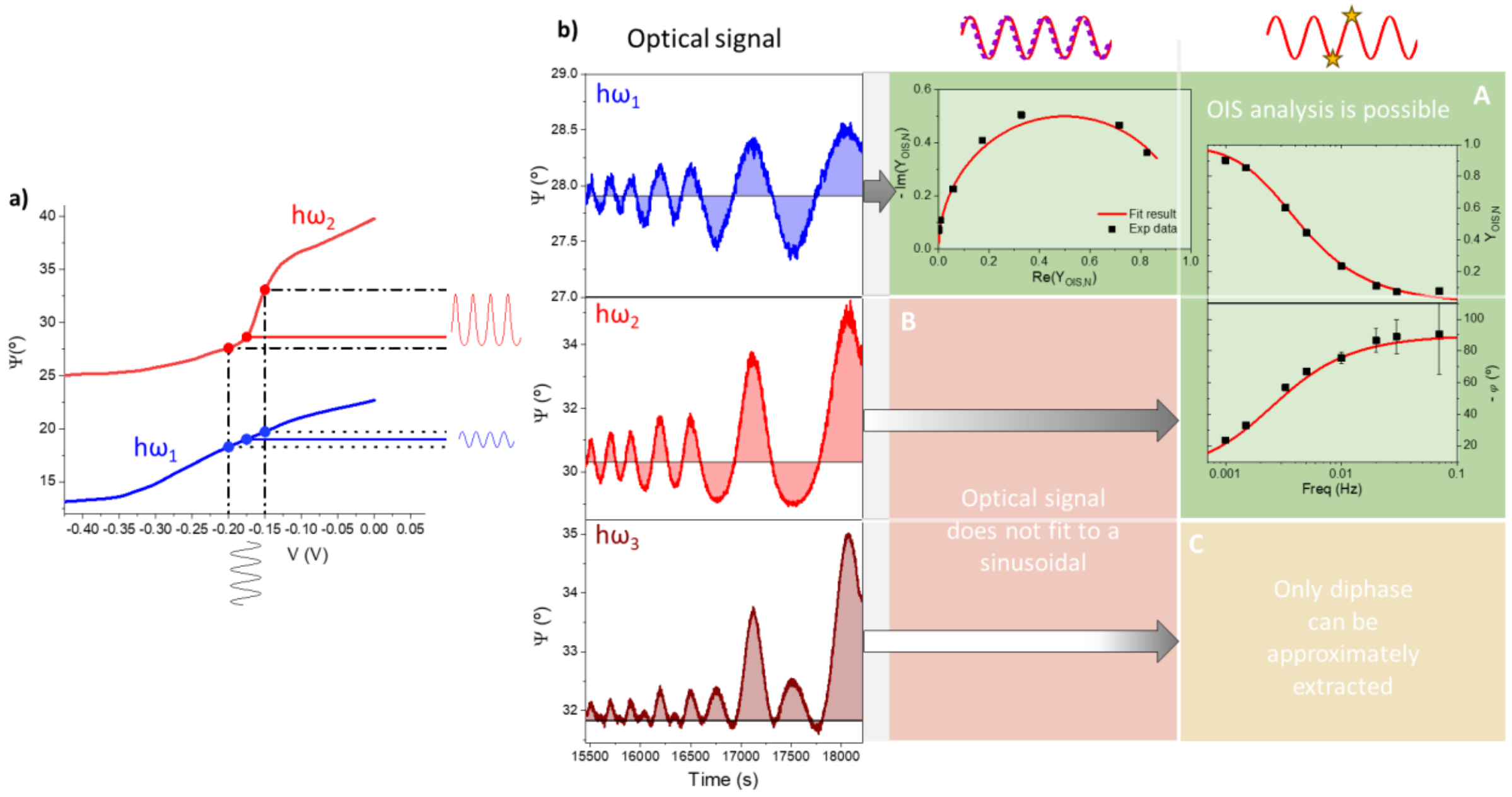

**Figure S1 a)** Input and output non-linear relation for two different photon energies ($h\omega$ = 2.564 eV, 2.956 eV marked in red and blue, respectively) and a representation of their approximately linear and non-linear expected response. **b)** Optoionic Impedance measurements simultaneously acquired at various photon energies ($h\omega$ = 2.564 eV, 2.956 eV, and 3.285 eV from top to bottom) and a table showing the different obtainable OIS possibilities depending on whether the analysis is done by sinusoidal fitting or manual point-extraction.

**S2 Equivalent circuit for OIS**

**S2.1 Transfer function for the insertion-limited case**

In order to clarify the expected behavior of charge transport in the device fabricated in the YSZ substrate, the most general expression for the EIS behavior should be taken into consideration. In this sense, the equivalent circuit derived by Jamnik and Maier[2,3] for dense MIEC thin films is considered to describe the functional layer of LSFx. **Figure S2-I** shows denotes the ionic and electronical pathways in purple and green, respectively. The variation of current through the cell in response to an AC voltage signal following the equivalent circuit of the system shown in **Figure S2-I** is composed, from right to left, by: (i.) a resistance $R_{CE}$ in parallel with a capacitance $C_{CE}$ due to oxygen insertion on the counter-electrode and the double layer formation;[4] (ii.) a resistance $R_{YSZ}$ in parallel with a capacitance $C_{CE}$ describing the pure ionic conduction through the YSZ and the capacitance coming from its dielectric nature, respectively; (iii.) bulk transport inside LSFx is described by $R_e$ and a capacitance $C_{Chem}$ represents the accumulation of ionic point defects in the LSF layer in response of a Nernstian voltage. Moreover, in the electronic line, electrical resistance of the film $R_e$ can also be considered. Finally, (iv.) the polarization resistance in the working electrode (WE) that can be neglected and the ionic insertion through the capping layer from the atmosphere. Also, an electrical capacitance in the LSF/YSZ interface ($Z_{e^-}$) blocks the electronic path from the YSZ. For a low electrically resistive film ($R_e \rightarrow 0$) the corresponding impedance for the (iii.) and (iv.) part is then described as:[2,3]

$$Z_{LSFx/atm} = \frac{R_i + Z_{atm} X \cdot \coth(X)}{Z_{atm} \cdot i\omega C_{chem} + X \cdot \coth(X)} \text{ , where } X = \sqrt{i\omega R_i C_{chem}} \qquad \text{(S1)}$$

Where $R_i$ indicates the ionic resistance of the film, $C_{chem}$ is the total chemical capacitance of the material, $\omega$ is the frequency of the applied AC voltage and $Z_{atm}$ is the impedance of the ionic insertion through the capping layer. A first approximation has already been done to this equation with respect to the one presented by Jamnik and Maier[2,3] as the LSF/YSZ interfacial capacitance has been considered to completely block the electrons to move through the YSZ ($Z_{e^-} \rightarrow$ inf., *i.e* $Z_{e^-} \rightarrow$inf.). Note that when in expression (S1), if the material could not accumulate charge, $C_{chem}$=0, then $Z_{LSFx/atm} = R_i + Z_{atm}$ i.e., the material would let all the oxygens flow through it with a resistance of $R_i$.

We know that the insertion of oxygen across the capping layer is practically zero as the $Al_2O_3$ layer blocks the equilibrium with the atmosphere. The equivalent circuit would be the one shown in **Figure S2-II** if $R_e$ was still considered and, $Z_{atm} \rightarrow \infty$. In our case, in which $R_e \rightarrow 0$, the circuit of **Figure S2-III** will make equation (1) look like

$$Z_{LSFx} = \frac{X \cdot \coth(X)}{i\omega C_{chem}} = \frac{R_i}{X} \coth(X) \tag{S2}$$

This expression is indeed the one given in the main text for an Open Warburg element, and is the one used to describe the expected behavior of the LSFx film during all the OIS experiments carried out in the study. As also explained in the main text, the frequency of study of the OIS analysis surpassed the characteristic time frequencies of the YSZ and CE processes allowing to treat them as a single resistance. In other words, as $\omega \gg \tau_{YSZ}, \tau_{CE}$, then $Z_{YSZ} + Z_{CE} \rightarrow R_{YSZ} + R_{CE} = R$. As explained in the main text, the $Y_N(\omega)$ was described by the product of the transfer function of the functional layer and the relation between the charge and applied voltage, $\frac{q_{LSFx}(\omega)}{V_{LSFx}(\omega)}$. For the circuit shown in **Figure S2-III** the transfer function results in:

$$H(\omega) = \frac{V_{LSFx}(\omega)}{V(\omega)} = \frac{I(\omega) \cdot Z_{LSFx}}{I(\omega) \cdot (Z_{LSFx} + R)} = \frac{1}{1 + R/Z_{LSFx}} \tag{S3}$$

If the studied film is no diffusion limited, then $R_i \rightarrow 0$ and the equivalent circuit becomes fully described by the ionic capacities placed in parallel with the electronic resistance of the interface**.** As a result, the equivalent circuit can be reorganized as in figure **Figure S2-IV** where the sum of capacities leads to the description of the film as a single capacity. Doing this approximation from equation **(S2)** one gets

$$Z_{LSFx} = \frac{1}{i\omega C_{chem}}; \quad H(\omega) = \frac{1}{1 + i\omega R C_{chem}} \tag{S4}$$

and so $Z_{LSFx}$ is the impedance term of a capacity and the transfer function is the one of an RC circuit.

## S2.2 Transfer function for the diffusion-limited case

For the diffusion limited case, simplifications must be done to recover **Equation (S3)** as the $H(\omega)$ of the element. If one takes the Randles circuit shown in **Figure 6c** in the main text, the most general expression of $H(\omega)$ that can be derived results in:

$$H(\omega) = \frac{V_{LSF}(\omega)}{V(\omega)} = \frac{I_{up}(\omega) \cdot Z_{LSF}}{I(\omega) \cdot Z_{Total}} = \frac{Z_{C_{dl}} \cdot Z_{LSF}}{R_{ely} \cdot (Z_{C_{dl}} + R_{inc} + Z_{LSF}) + (R_{inc} + Z_{LSF}) Z_{C_{dl}}} \tag{S5}$$

Where $\mathrm{R}_{ely}$ is the resistance of the electrolyte, $\mathrm{R}_{inc}$ the insertion resistance and $\mathrm{Z}_{\mathrm{C}_{dl}}$is the double-layer capacity element. As explained in the main text, $\mathrm{R}_{ely} \ll \mathrm{R}_{inc}$. As a first approximation, $\mathrm{R}_{ely}$ can be neglected from this equation. If this is considered, then $H(\omega)$ is described again by **Equation (S3)**.

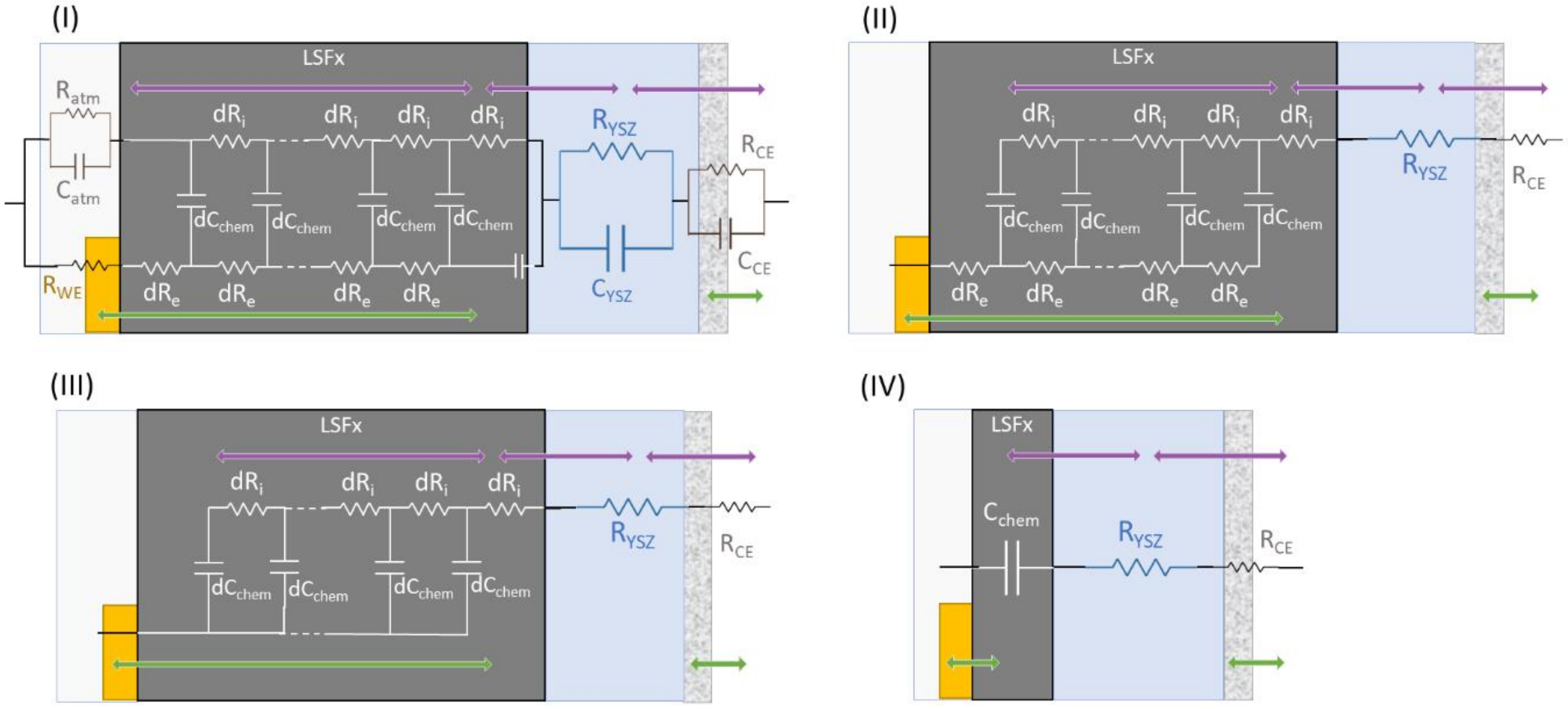

**Figure S2**. Simplification of the equivalent circuit of the sample under study. Purple arrows indicate the ionic motion paths ($O^{2-}$ paths) and green arrows the electronic paths.

### S2.3 Derivation of $\frac{q_{mat}(\omega)}{V_{mat}(\omega)}$

The second important term to solve in order to get the OIS expression that fits each model system is the $\frac{q_{mat}(\omega)}{V_{mat}(\omega)}$ relation. To understand its real meaning of this term we can start from the simplest possible model that can describe our material: the capacitor. A capacitor element is responsible for accumulating charge $q_{mat}$ when applying a voltage drop at its ends, $V_{mat}$. In the case of MIECs this charge is in the form of charged ions being incorporated into the layer. The relation of accumulated charge and voltage in a pure capacitor is simply $C$.

Now, in the case of the most general expression of a MIEC, considering both electrical and ionic resistivities, the material could be described with the transmission line (circuit of **Figure S2-II** excluding the incsertion resistance terms). In this case, the total accumulated charge is distributed along the different capacities of the line, that can be seen as the different charging state of different points along the thickness of the material. The closer to the ionic insertion, the faster the capacity will charge, as ions must cross the different $dR_i$ in order to reach the inner part of the material.

Electrically, the state of charge of each capacitor of the transmission line would be related to the voltage drop at its ends as $dq(x) = dC \cdot (V_i(x) - V_e(x))$, where $V_i(x)$ and $V_e(x)$ are the voltage in the ionic and electronic path, respectively. Nevertheless, OIS does not care about the electrical contribution to the charging of the capacitors, so the state of ionic charging of the capacitor is indeed:

$$dq(x) = dC \cdot V_i(x) \tag{S6}$$

Where $dC = \frac{C}{L}dx$, being $C$ the total capacity of the film. So, to obtain the total amount of charge in the material, this expression must be integrated for all the length $L$ of the material.

In order to get the expression for $V_i(x)$, several calculations must be done involving the resolution of the following differential equation system.

| **Impedance definitions** | **Kirchoff and Ohm's laws** | **Boundary conditions** |
|---|---|---|
| $dR_i = \frac{R_i}{L}dx$ | $\frac{dV_i(x)}{dx} = -\frac{R_i}{L}I_i(x)$ | $V_i(0) = V_{mat}(\omega)$ |
| $dR_e = \frac{R_e}{L}dx$ | $\frac{dV_e(x)}{dx} = -\frac{R_e}{L}I_e(x)$ | |
| $dZ_C = \frac{Z_C}{dx}L$ | $-\frac{dI_i(x)}{dx} = \frac{dI_e(x)}{dx} = \frac{V_i(x) - V_e(x)}{Z_C L}$ | $I_i(L) = 0$ |

Where $R_i$ and $R_e$ are the total ionic and electrical resistance of the material; $V_i(x)$ and $V_e(x)$ are the voltage at the different points of the ionic and electronic path, respectively; and $I_i(x)$ and $I_e(x)$ are the position dependent current at the ionic and electronic paths, respectively. The resulting expression for $V_i(x)$ ends up being:

$$V_i(x) = V_{mat}(\omega)\left(\frac{\frac{R_e}{R_i}\frac{sinh(k[x-L])}{kL} + cosh(k[x-L]) + R_e\frac{x}{L}}{R_i\, cosh(kL) + \frac{R_e}{kL}sinh(kL)}\right) \quad where\ k = \frac{1}{L}\sqrt{i\omega(R_i + R_e)C} \tag{S7}$$

From which the most general $\frac{q_{mat}(\omega)}{V_{mat}(\omega)}$ expression is obtained as:

$$\frac{q_{mat}(\omega)}{V_{mat}(\omega)} = \frac{\int_0^L C/L \cdot V_i(x)dx}{V_{mat}(\omega)} = \frac{C}{R_i\, cosh(kL) + \frac{R_e}{kL}sinh(kL)} \cdot \left(\frac{R_i sinh(kL)}{kL} + R_e\frac{(cosh(kL)-1)}{(kL)^2} + \frac{R_e}{2}\right) \tag{S8}$$

And the most general expression for the impedance of the element can be calculated as:

$$Z_{mat} = \frac{V_{applied}}{I_i(0)} = \frac{R_i\, cosh(kL) + \frac{R_e}{kL}sinh(kL)}{kL\, sinh(kL) + \frac{R_e}{R_i}(cosh(kL)-1)} \tag{S9}$$

Notably, the expressions presented in **Equations (S8) and (S9)** can be simplified to the Open Warburg when $R_e \rightarrow 0$ and to a capacitor when $R_i \rightarrow 0$. The resumé of all the possible $\frac{q_{mat}(\omega)}{V_{mat}(\omega)}$ and $Z_{mat}$ are summarized in **Table S1**.

**Table S1** Impedance element and $\frac{q_{mat}(\omega)}{V_{Lmat}(\omega)}$ for any MIEC in the different limiting cases

| **Element** | $\boldsymbol{Z_{mat}}$ | $\boldsymbol{\frac{q_{mat}(\omega)}{V_{Lmat}(\omega)}}$ |
|---|---|---|
| General MIEC | $\frac{R_i\, cosh(kL) + \frac{R_e}{kL} sinh(kL)}{kL\, sinh(kL) + \frac{R_e}{R_i}(cosh(kL) - 1)}$ | $C \cdot \frac{R_i \frac{sinh(kL)}{kL} + R_e \frac{cosh(kL) - 1}{(kL)^2} + \frac{R_e}{2}}{R_i\, cosh(kL) + \frac{R_e}{kL} sinh(kL)}$ |
| $R_e \rightarrow 0$ | $\frac{R_i}{kL} \coth(kL)$ | $C \cdot \frac{\tanh(kL)}{kL}$ |
| $R_e, R_i \rightarrow 0$ | $\frac{1}{i\omega C}$ | $C$ |

The difference in the OIS behavior of a system composed by an insertion resistance $R$ and an Open Warburg element with ionic resistance $R_i$ is presented if **Figure S3**. The same figure also includes the dependance of the response to a different total capacity of the film. Interestingly, the shape of the amplitude and diphase graphs depends on the ion motion part, i.e., $R_i$ and $R$. Whilst the frequency at which each feature happens depends on the product of $C$ and the two characteristic resistances of the system.

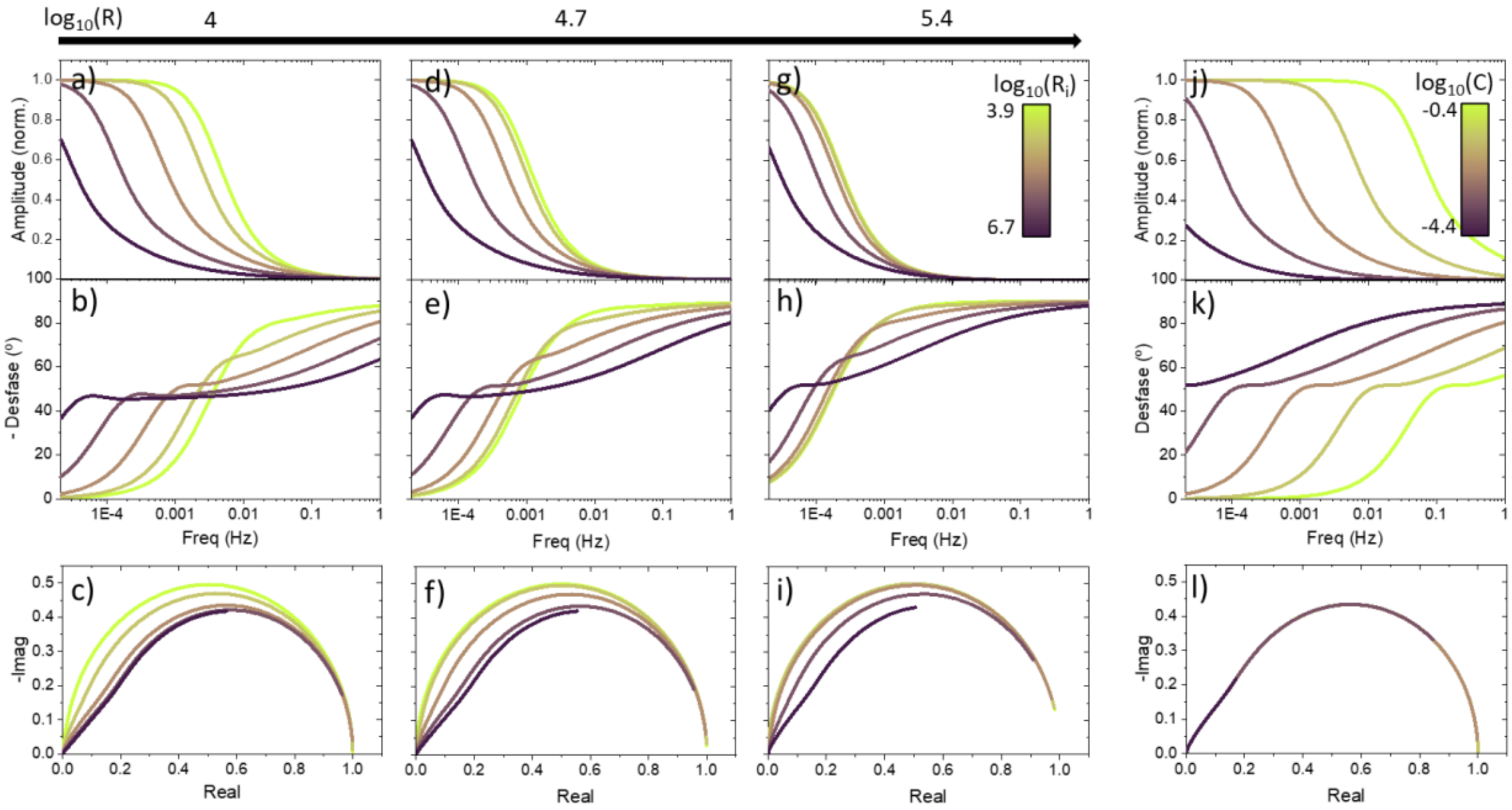

**Figure S3**. **a-i)** Bode and Nyquist OIS Admittance plots for different R and Ri values calculated for a constant C=0.004F. **j-l)** Bode and Nyquist OIS Admittance plots for different C values with constant R=10kΩ and Ri=200kΩ values.

**S2.4 Verification of the diffusion-limited model implementation**

Simulations of the insertion-limited sample can be done taking the $R_{YSZ}$ corresponding to the measurement temperature and different values for the oxygen diffusion. **Figure S4a and S4b** shows the system's behavior when the ionic diffusion of the film is decreased (increase of $R_i$) and are compared with the experimental data from the LSF sample measured at $E_1$= -0.075V. This voltage was chosen for the comparison due to the inverse relation of oxygen mobility and vacancy concentration (lower cathodic potential will generate more vacancies). It is clearly seen that as the diffusivity is lowered and becomes limiting, the response at higher frequencies starts to change its shape, as the defects do not have time to charge the film uniformly and only the first layers of material are charged. Nevertheless, when the frequency is low enough, the defects have more time to move across the film and to equilibrate it with the applied bias making the spectra coincide.

The observed behavior is well described by the low ionic resistivity regime as the spectra follows the semicircle in the Nyquist plot and does not show a visible decrease in the diphase when increasing the frequency. The low thickness of the LSFx films leads to a situation in which $R_i$ is orders of magnitude lower than the $R_{YSZ}$. In the studied range of frequencies, diffusion in OIS can start to be visible when Ri>1000 Ohm, which corresponds to a $D^{q}_{o,lim}$=2e-13 cm$^2$/s. Equivalently, we can obtain the vacancy diffusion $D_v$ by simply applying

$$D^{q}_{o} = f n_v D_v \qquad \text{(S10)}$$

Where f=0.69 is the correlation factor,[5,6] $n_v$ the site fraction of oxygen vacancies $\frac{[V_O^{\cdot\cdot}]}{[O_O^{x}]} = \frac{\delta}{3-\delta}$. If a $\delta = 0.001$ is considered looking at the defect chemistry extracted for LSF50 in this paper, then $D_{v,lim}$=2.9e-10 cm$^2$/s. From reference,[6] one can extrapolate the vacancy diffusion $D_v$ for this material and get a value of $D_v$=4.8e-9 cm$^2$/s. As the number of vacancies will increase as the material gets more reduced, this value can be considered as the lowest oxygen diffusion the LSF will have in the experiment conditions. This confirms again why the observed behavior in the experimental data from **Figure S4a and S4b** could be well fitted with a capacity element describing the layer and confirms the initial hypothesis of non-diffusion limitations in the measured response.

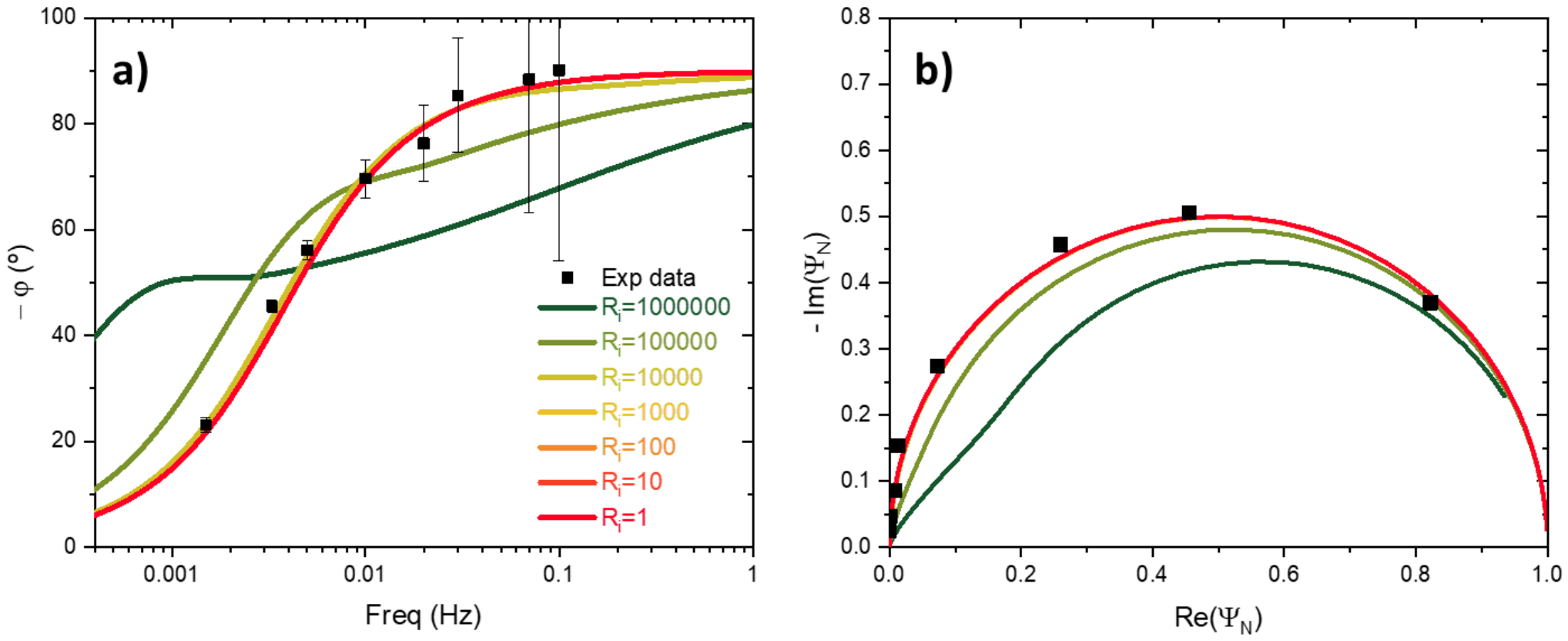

**Figure S4**. OIS simulations for different functional film ionic resistances in the insertion limited case, with a constant $R_{YSZ}$. Figure **a)** shows the diphase for different frequencies and Figure **b)** the Nyquist Admittance OIS plot, respectively. Black dots in a-b show the experimental OIS data measured for $E_1$=-0.075V

## S3 Discussion on the OIS applicability and limitations

After the validation of the technique and the exploration of it in different usage cases, it is important to present an overview of the presented OIS technique and address several possible concerns. These include considerations such as the temporal and optical resolution, as well as the yet commented optical signal linearity with the external input and. The first issue comes from optical spectra acquisition time. The acquisition time should be small compared with the specific time response of the system. If the average time of the reading is on the same order as the change in optical properties, the technique will not be sufficiently accurate to determine their changes correctly. Differently from single measurements in which the optical response can be measured at each energy separately with long averaging times, in-situ measurement takes place continuously and each spectrum is acquired in less time. Single measurements can regulate the intensity of the laser incising the sample at each of the energies measured to get the optimal output signal to be detected by the sensor. In the case of in-situ measurements, this optimization of light intensity is done at the beginning of the experiment and remains constant all along the measurement. This procedure must also be executed with care, as a substantial alteration in optical characteristics, particularly a substantial increase in absorption, has the potential to significantly diminish the detected signal, possibly to the point of signal loss. The initial calibration of light intensity should ideally be conducted when the system is in a state where the output signal is at its minimum, allowing for the adjustment of laser intensity accordingly. This strategic procedure guarantees a stronger signal and, consequently, enhanced sensitivity throughout the measurement.

Parallel to other Impedance techniques, signal linearity only comes if the applied voltage amplitude ($E_0$) is low enough. The main text discusses that this effect relates to the linearity with the optical signal, but applying higher voltages can also lead to an error in the characterization of the material. In the example presented in this study consisting in $C_{chem}$ characterization one can see that there is an important $C_{chem}$ vs V dependance. In this sense, the $C_{chem}$ measured at a certain $E_1$, which can be denoted as "equivalent chemical capacitance" $C_{\text{chem,eq}}$ is dependent on the amplitude of the applied bias as

$$C_{\text{chem,eq}}(E_1) = \frac{\int_{E_1-E_0}^{E_1+E_0} C_{chem}(V)dV}{\int_{E_1-E_0}^{E_1+E_0} dV} \quad \text{(S7)}$$

It is evident from the formula that the lower the applied $E_0$, the near the value will be to the real Cchem as $\lim_{E_0 \to 0} C_{\text{chem,eq}}(E_1) = C_{\text{chem}}(E_1)$. So, the chosen $E_0$ must find a good agreement between sufficient optical signal amplitude to perform the analysis and low enough voltage amplitude to have a precise $C_{chem}$ characterization.

**S.4 OIS setup and microstructural analysis of the sample**

**Figure S5a** shows a schematic drawing of the sample's structure. Since the gold contact was not deposited over the entire functional LSFx layer, a window with sufficient size for the ellipsometer spot remained open for OIS measurements. The silver paste on the opposite side of the YSZ substrate allowed the application of voltages across the solid electrolyte, promoting the entry or exit of oxygen into the functional layer through the YSZ, as indicated in purple in **Figure S5a**. An image of the setup is displayed in **Figure S5b**. The *Linkam THMS600* station was positioned at the center of the ellipsometer setup plate to adjust the height, ensuring that the ellipsometer beam hit the center of the sample. Once this adjustment was completed, to maintain a stable temperature, a laboratory-made lid with openings for ellipsometer light passage was placed on top. The inset in **Figure S5b** reveals the *Linkam THMS600* chamber, showing the hot plate for temperature control and the established connections on the working and counter electrodes of the sample.

Prior to the measurements, thin film was structurally characterized. The X-ray diffraction (XRD) patterns of the as-prepared samples are presented in **Figure S5c**. In all the samples, the functional thin film material (LSF50, SFO or both) displays a pseudo-cubic polycrystalline structure with a (h00) preferred orientation. The measured cell parameter agrees with the experimental references.[7–10] The epitaxial growth of Al2O3 on top of the sample is also confirmed by the α-110 peak at around 38º. Also, atomic force microscopy (AFM) measurements of surface topography of the functional layers were performed prior to the $Al_2O_3$ capping deposition (**Figure S5**) confirms the epitaxially and good structural order of the as grown layers as foretold by the XRD.

Furthermore, Atomic Force Microscopy (AFM) measurements of the surface of each sample prior to $Al_2O_3$ deposition are provided (**Figures S5e-f**). The observed low surface roughness and uniformity across the layer surface confirm the epitaxial nature and well-organized structure of the as-grown layers. AFM measurements of the alumina layer's surface also corroborate its low roughness and orderly growth, aligning with the XRD analysis results. A porous layer or a non-conformality of the capping could have led to deviations in the measurements with respect to the ideal capping as the layer would be constantly addressing equilibrium with the atmosphere.[11,12] After the study, a cross-section SEM image of the samples was taken to confirm the dense and epitaxial growth observed in the XRD and AFM characterization. **Figure S5d** shows the SEM image of the LSF50 sample where each of the components are indicated with different color codes.

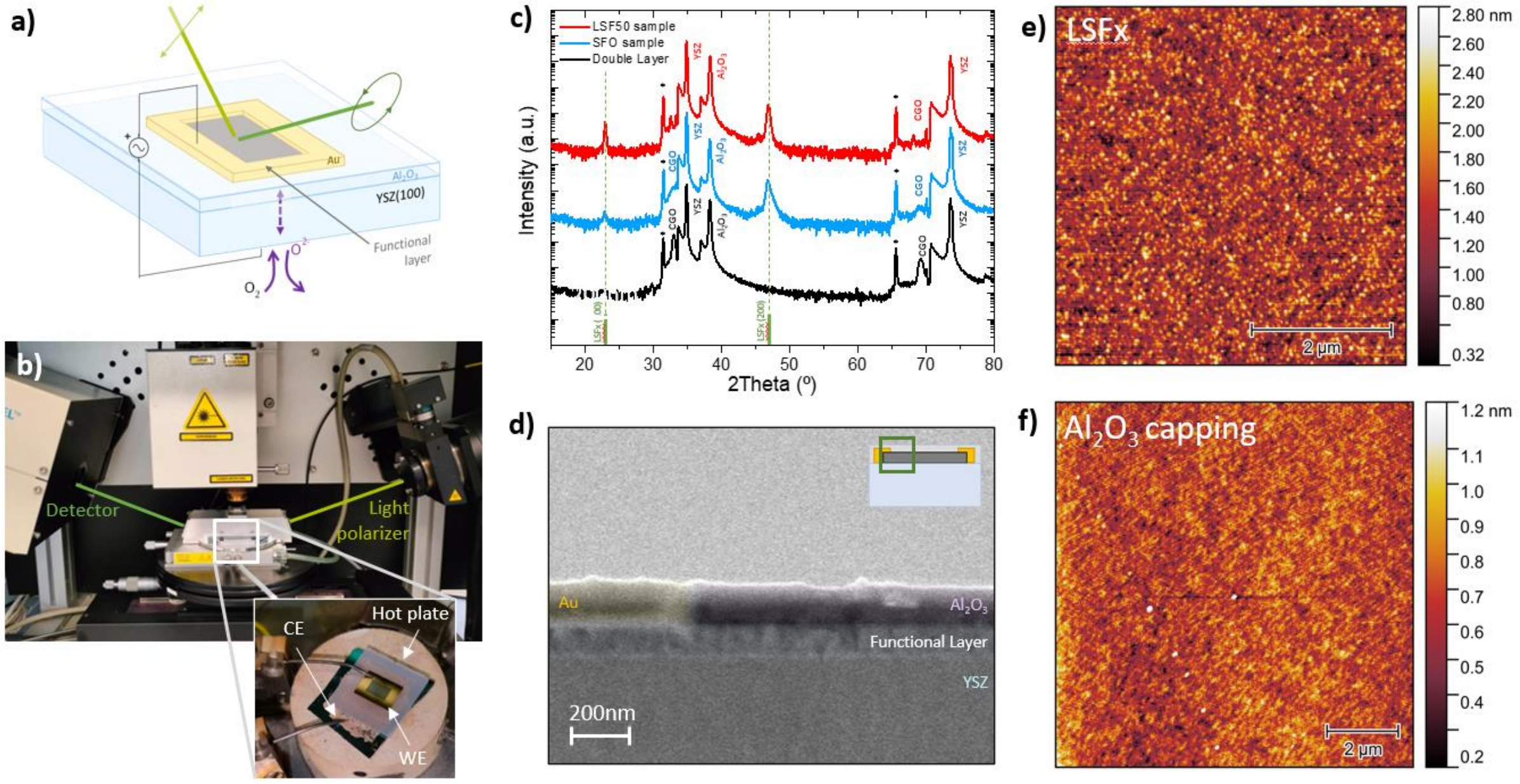

**Figure S5** a) Schematic representation of the samples under study. b) Image of the in-situ ellipsometry setup. c) XRD pattern of the samples. d) SEM cross section of one of the measured samples. f) and g) AFM images of the LSF50 surface and the Al2O3 capping, respectively.

**S5 Accuracy of the technique**

These samples provided a straightforward model to demonstrate the applicability of OIS, allowing us to establish an equivalent circuit for analysis. The circuit utilized for OIS was essentially a streamlined version of the electrical equivalent model. Since current measurement is performed simultaneously during the experiment, it can also be utilized for Electrochemical Impedance Spectroscopy (EIS) analysis. **Figures S6b, S6c and S6d** depict the resulting Bode and Nyquist plots obtained from an EIS measurement. These plots encompass the three characteristics of the processes discussed in the paper. Upon closer examination of the Nyquist plot (**Figure S6c**), one can observe two semicircles closing at 30kΩ and 40kΩ, respectively. These semicircles correspond to the anticipated response of a process whose equivalent components are a resistance and a capacitance arranged in parallel. Indeed, these correspond to the previously mentioned processes (ii) and (i), respectively. The point at which the first semicircle closes correspond to the resistance of YSZ, while the 40kΩ value corresponds to the total system resistance $R_{YSZ}$+Rs. The third process involved in the response of the sample was the accumulation of charges in the functional layer.

The third process involved in the sample's response is the accumulation of charges within the functional layer. This process also leaves its mark on the Nyquist plot, manifested as a sudden and nearly vertical increase. The rationale behind this response becomes evident when analyzing the Bode plot (**Figure S6b**). By examining the sequence starting from higher frequencies and moving towards lower frequencies, a noticeable pattern emerges: the phase shift, in terms of absolute value, decreases, while the amplitude increases. This behavior occurs as the system surpasses the time constant (τ) associated with the first two processes. Once this point is exceeded, the phase shift begins to increase again (still in terms of absolute value), along with the amplitude. This phenomenon is the one caused by film charging. This occurrence can be explained simply by considering that as we approach a DC measurement, the current tends to zero i.e., $Z(\omega) = \frac{V(\omega)}{I(\omega)}$ tends to infinity for low frequencies. With more time available to achieve equilibrium, the current gradually shifts out of phase with the applied voltage.

By employing the appropriate equivalent circuit for EIS, it becomes possible to calculate the capacitance value of the film and, consequently, the $C_{chem}$ value. **Figure S6a** presents a comparative analysis of $C_{chem}$ values for the sample featuring a 99 nm LSF50 functional layer measured with both techniques. Notably, the calculated Cchem values from each method closely align, falling within the expected margin of error. This consistency reinforces the conclusions drawn from OIS analysis, affirming the method's reliability and applicability.

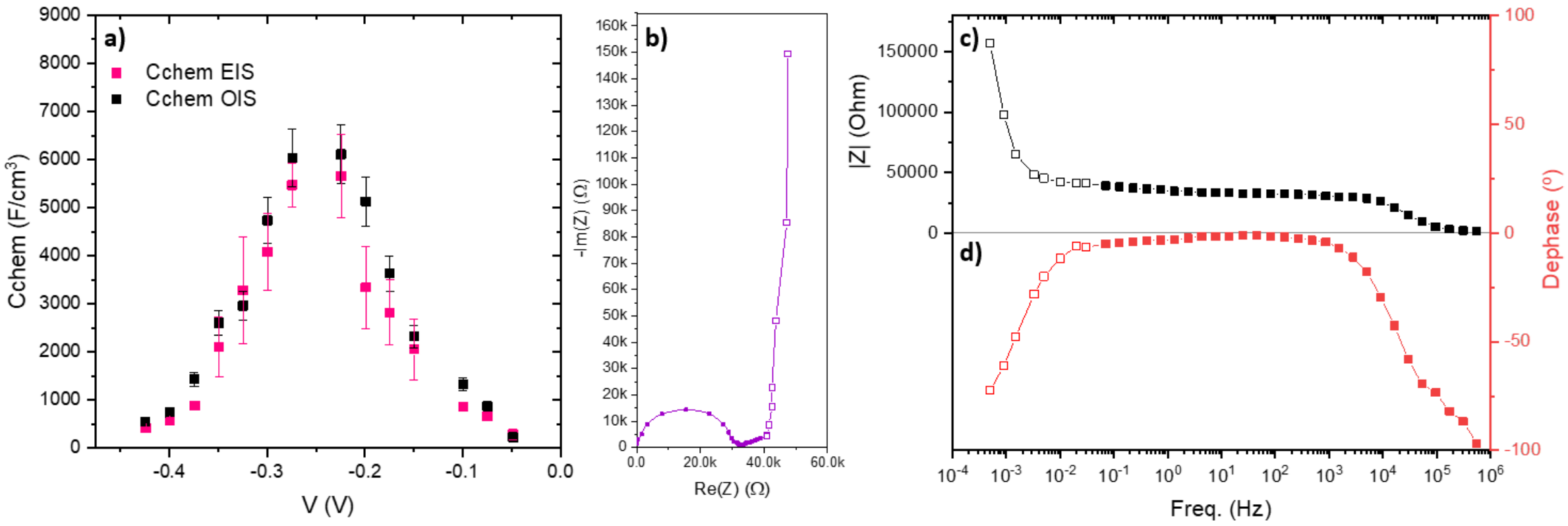

**Figure S6 a)** Comparison of the results for the Cchem obtained with EIS and OIS respectively, showing good agreement between both measurements. **b)** Nyquist EIS for one of the applied $E_1$ and **c-d)** Bode plot of the same data. Colored squares indicate the EIS measurement taken with the *Biologic SP-150* and white filled squares the ones extracted from simultaneous current measurement during OIS.

**S6 Device-like optical response optimization using OIS**

If we consider the insertion-limited sample as a device in which switching of material properties is essential (e.g., conductivity, magnetic characteristics or optical properties), OIS can be employed to investigate the optimization of the material's response to external stimuli. Here we consider an electrochromic device based on LSF50 thin films deposited on YSZ, as the system studied in section 3.1.1, whose objective is to change its transmittance as much and as fast possible. In the case of the studied sample, this translates to a device with a switching time defined as $\tau$=RC. To reduce the device's response time, there are two possible strategies: the reduction of R, entailing a decrease in the resistance to oxygen insertion within the film, or the reduction of C, which represents the equivalent capacitance of the layer. The reduction of the layer's capacity implies reducing the dimensions of the functional layer, thereby influencing properties dependent on this parameter, such as conductance. As a matter of example, we have considered a LSF50 single layer and we have proceeded to reduce the insertion resistance by reducing the YSZ substrate thickness, allowing measurements to be taken for three different values of R. The analysis of OIS measured at the potential $E_1$ of the greatest change, i.e. the maximum of $C_{chem}$, is depicted in **Figure S7a** for the three measured resistances. A shift in the response profile is evident, resulting in its occurrence at elevated frequencies as resistance is decreased. In other words, for the same frequency now the material presents a larger variation of oxygen concentration.

These results lead to a rather straightforward conclusion, as a fast response will occur when the insertion resistance is minimized. Therefore, the enhancement of the device would necessitate minimizing the YSZ layer's thickness, effectively orienting the device's design towards an all-thin-film-based configuration. Interestingly, in this case oxygen diffusivity through the layer may not be negligible and the model developed in **Supplementary Information Section S2** may be required to properly interpret OIS.

Once the voltage where the most significant change in oxidation state occurs is identified, it becomes feasible to study the response of device characteristics, such as transmittance (T). The technique's capability to analyze both $\Delta$ and $\Psi$ components for the desired energy allows for the extraction of the T value by considering the sample as a semi-infinite homogeneous material.[13] **Figure S7b** is dedicated to the examination of this particular property within the red-light spectrum, specifically at a photon energy of 1.968 eV ($\lambda$=630nm). One compelling reason for focusing this spectral range lies in the material's pronounced absorbance change as its oxidation changes i.e., as the [$Fe^{4+}$] changes.[10] At this stage, it is essential to establish a key parameter. The optical switching contrast $\gamma_o$,[14] comes to the forefront as the parameter of interest for this analysis and it is represented by the relative variation of transmittance:

$$\gamma_o = \frac{T - T_{ref}}{T_{ref}} \qquad (13)$$

where $T_{ref}$ refers to the initial transmittance and T to the transmittance when the AC voltage is applied. Four different voltage amplitudes ($E_1$= 25mV, 50mV, 100mV and 200mV) were applied with $E_0$=-225mV at the device with the smallest insertion resistance. Given the non-linearity of T changes in response to voltage, particularly in the case of substantial voltage differences, fitting the measured transmittance as a sinusoidal curve as done for Ψ in equation (2) is not applicable. Also, the capacity can no longer be considered as constant as the material is moving through its different oxidation states, especially if the difference between the applied voltage and $E_1$ increase, not only the behavior of optical properties does not follow a sinusoidal relation, but current will also deviate from an ideal periodic oscillation. Nevertheless, one can note the point of maximum change in the material's properties and extract from it the transmittance value.

As $\gamma_o$ will not give the same value for the maximum value of transmittance $T_{max}$ and the minimum $T_{min}$, a novel parameter can be derived to standardize the variation in transmittance relative to a reference value depending on the frequency. This parameter is the dynamic optical switching contrast, $\gamma_{o,AC}$:

$$\gamma_{o,AC} = \frac{T_{max} - T_{min}}{T_{min}} \qquad (14)$$

Where $\gamma_{o,max}$ and $\gamma_{o,min}$ are the optical switching contrast for the maximal and minimal transmittance measured at each frequency, respectively.

In this sense, **Figure S7b** shows the normalized change in transmittance with the $\gamma_{o,AC}$ parameter showing that the change is higher when moving to lower frequencies, giving the device more time to change. Notably, for a given change in transmittance, one can observe that the same effect in the optical properties can be achieved within frequencies of several orders of magnitude difference depending on the applied voltage.

In this section, the use of OIS technique has been examined to investigate the response of a device. Initially, the response speed of the device at different frequencies was assessed, and a strategy to optimize its performance was sought. Conclusions drawn from these measurements suggest that employing a thin-layer electrolyte would considerably improve the efficiency of the switching. In a subsequent phase, the capabilities of in-situ ellipsometry, combined with the OIS technique, were leveraged to compute alterations in transmittance using a model. The latest study not only enabled the measurement of the speed of change but also the specific value of the observed alteration in one of the most crucial properties in electrochemical and photonic devices.

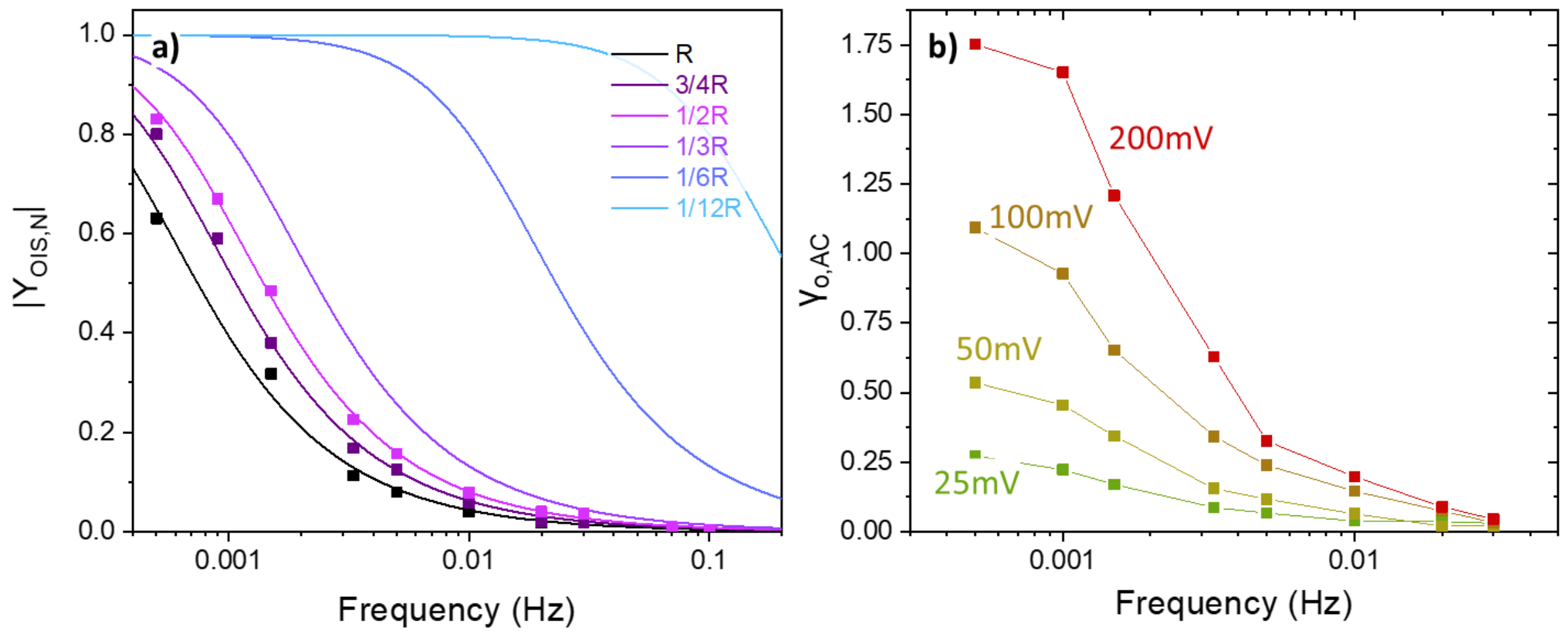

**Figure S7.** a) Bode plot for three different R values: R, 2R/3 and R/2 and the extrapolation of the expected response for lower insertion resistances. b) Dynamic optical switching contrast measured for different voltage amplitudes and frequencies

## S7 Diffusion-limited sample electrochemical characterization

Before initializing the OIS experiments, the optimal $E_1$ had to be chosen. For this purpose, cyclic voltammetry (CV) measurements were performed. A cycling speed of 0.5mV/s was decided, and the resulting CV loop is presented in **Figure S8a**. The peaks for the oxygen incorporation and disincorporation can be clearly seen at -125mV and -225mV, respectively, consistent with the results reported in literature.[15] For this reason, OIS spectra were chosen to be measured at an $E_1$=-0.2V. **Figure S8b and S8c** show the EIS measurement results performed before launching the OIS experiment. The spectrum could be fitted with the Randles circuit obtaining a $R_{ely}$=39Ω, a $C_{dl}$=9 μF, and a $R_{inc}$=10kΩ.

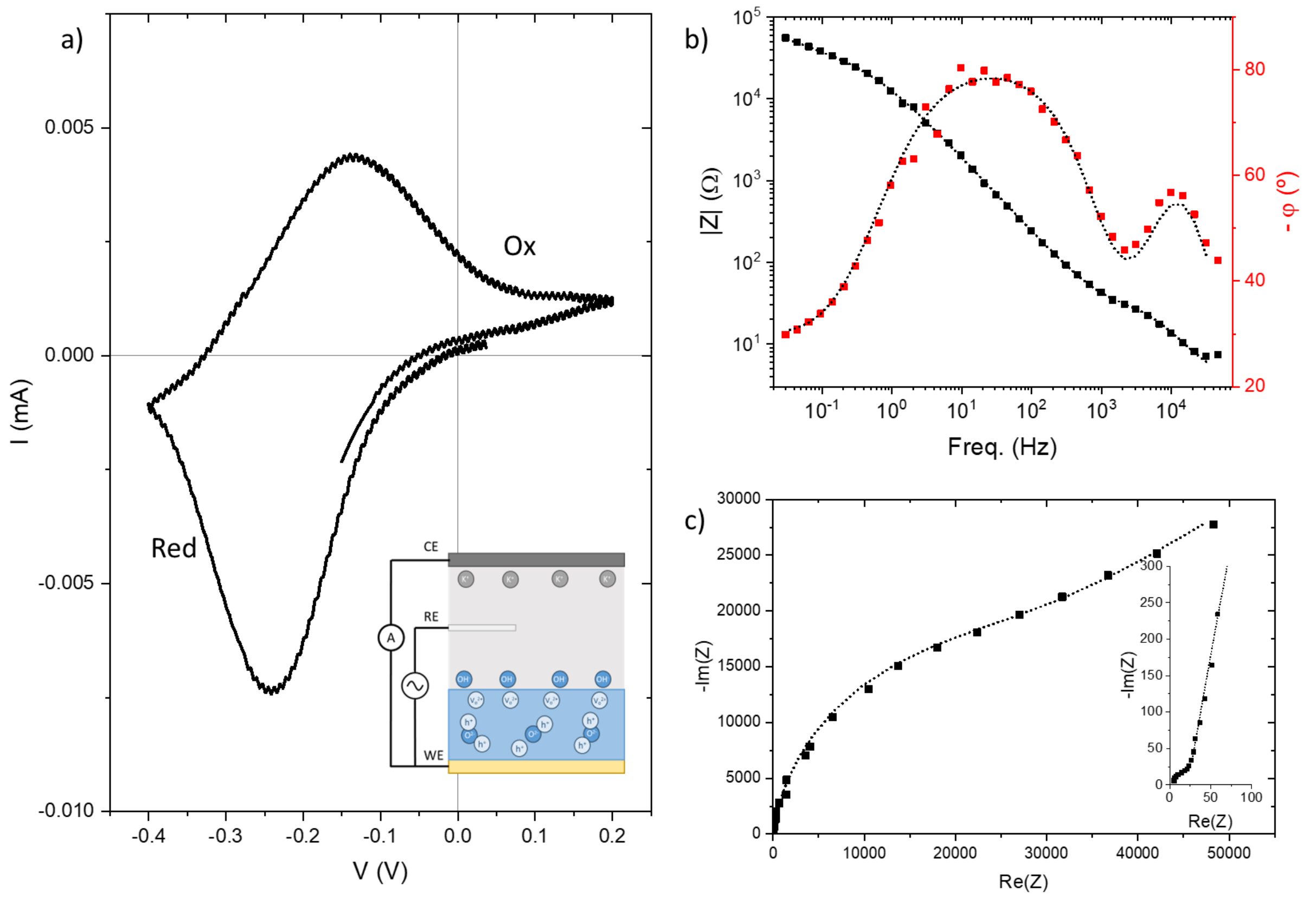

**Figure S8 a)** Cyclic Voltammetry of the diffusion-limited system measured at a cycling speed of 0.5mV/s. **b)** EIS Bode plot measured at -0.2V and the corresponding Nyquist plot **(c)** with an inset with a zoom for the high frequency spectra. Dashed data in b) and c) corresponds to the fitting with the equivalent circuit.